\documentclass[11pt]{article}
\usepackage[centertags]{amsmath}
\usepackage{amsfonts}
\usepackage{newlfont}
\usepackage{graphicx}
\usepackage{eurosym}
\usepackage{epstopdf}
\usepackage{amsmath,amssymb, xcolor}

\newtheorem{proposition}{Proposition}[section]

\newtheorem{lemma}{Lemma}[section]

\newtheorem{rem}{Remark}[section]
\newcommand{\F}{\cal{F}}

\newcommand{\G}{\cal{G}}

\newcommand{\Fs}{{\cal{F}}_s}
\newcommand{\Fu}{{\cal{F}}_u}
\newcommand{\FT}{{\cal{F}}_T}

\newcommand{\Ft}{{\cal{F}}_t}
\newcommand{\Gs}{{\cal{G}}_s}

\newcommand{\Gt}{{\cal{G}}_t}
\newcommand{\Ht}{{\cal{H}}_t}

\newcommand{\E}{\mathbf{E}}

\newcommand{\e}{\mathrm{e}}

\newcommand{\lam}{\lambda}

\def\ds{\displaystyle}

\def\cF{{\cal F}}

\def\cL{{\cal L}}

\def\cN{{\cal N}}

\def\tsigma{{\tilde\sigma}}

\def\ds{{\partial_s}}

\begin{document}
\title{CVA and vulnerable options in stochastic  volatility models}

\author{E. Alos\thanks{Dept. of Economics and Business - University Pompeu Fabra, \texttt{elisa.alos@upf.edu}}, F. Antonelli\thanks{DISIM - University of L'Aquila, \texttt{fabio.antonelli@univaq.it}}, A. Ramponi\thanks{Dept. of Economics and Finance - University of Rome Tor Vergata, \texttt{alessandro.ramponi@uniroma2.it}}, S. Scarlatti\thanks{Dept. of  Enterprise Engineering - University of Rome Tor Vergata, \texttt{sergio.scarlattii@uniroma2.it}}}


\maketitle


\begin{abstract}
In this work we want to provide a general principle to evaluate the CVA (Credit Value Adjustment) for a vulnerable option, that is an option  subject to some default event, concerning the solvability of the issuer. CVA is needed to evaluate correctly the contract and it  is particularly important in presence of WWR (Wrong Way Risk), when a credit deterioration determines an increase of the claim's price. In particular,  we are interested in evaluating the CVA in  stochastic volatility models for the underlying's price (which often fit quite well the market's prices) when admitting correlation with the default event. By cunningly using Ito's calculus, we provide a general  representation formula applicable to some popular models  such as SABR, Hull \& White and Heston, which  explicitly shows the correction in CVA due to  the processes correlation.  Later, we specialize this formula and construct its approximation for the three selected models. Lastly, we run a numerical study to test the formula's accuracy, comparing our results with Monte Carlo simulations.

\medskip




\end{abstract}

\section{Introduction}

\label{intro}

Defaultable claims are derivatives that are subject to some default event, which  concerns the solvability of the counterparty  before the final settlement of the transaction.  This is the so called "Counterparty Credit Risk" (CCR), an immediate consequence of it being  that the product's price needs an adjustment to include in its quotation the possibility of default. This adjustment, which results in a  price reduction, is commonly known as {\bf Credit Value Adjustment }(CVA) and it was first introduced in a paper by Zhu and Pykhtin (\cite{ZP07}).
The last financial crisis (2007-2008) has greatly increased the monitoring and pricing of CCR on OTC-markets products and  many researchers and practitioners tried to develop a general framework for a better assessment of the CVA evaluation to compensate a derivatives holder for taking CCR. Indeed, along the years, other value adjustments have been additionally considered leading to the acronym (X)VA.  An updated overview of the recent research directions under investigation is presented in \cite{GSZ17}.

Many works concentrate on CVA evaluation for swaps while others focus on European options. In this case , when the risk relates only to the issuer, these contracts are called vulnerable options and one can find a vast literature  under this denomination (see e.g. \cite{Fard15}, \cite{BV18},\cite{BRH18}, \cite{ARS2} and references therein, the last three focusing more on CVA).

Typically the default event is characterized by means of a random time, representing the time of default. At the time of default there might be a total loss for the investor or a partial  recovery of the investment's current value might be possible.

The difficulty in the evaluation of the CVA is twofold.  First of all the default time might be not completely measurable with respect to the information generated by the market prices, since it reflects also other exogenous factors, secondly even under full knowledge of this random time, the derivative's evaluation will call  for the  joint distribution of the random time and the price processes, usually very difficult to know.

In this framework we consider the CVA evaluation of a vulnerable European option under a variety of stochastic volatility models. The importance of this choice lies in the fact that stochastic volatility models better fit the market, reproducing the smiles and skews of the implied volatility.
Pricing of vulnerable options under the Heston model was  already  discussed in  some papers ( \cite{LYK16} and \cite{WWZ17}) under  the structural default framework (see \cite{Kl96}).

Alternatively, one might use the so called intensity approach (introduced in \cite{Lando98} to price defaultable bonds) in order to characterize the distribution of the default time conditionally to the information generated by the market prices. In a stochastic volatility setting a first result was obtained in \cite{W17} in discrete time, assuming  a GARCH model for the underlying coupled with default intensity, but the literature is scarce in continuous time.

In the present paper,  we describe the joint dynamics of the asset prices, of the default time and of the other stochastic factors as a Markovian system whose components may exhibit correlation. We remark that the presence of correlation among processes is crucial, indeed under independence between the intensity and the price/volatility processes, the evaluation formula reduces to two separate evaluations: the classical default-free derivative price multiplied by a   factor (similar to a bond  price paying $1$ at maturity) induced by the intensity process.

When in presence of correlation, computations do not come out as easily and  we employ a keen integration by parts formula, inspired by  a  technique developed by E. Alos (\cite{A}), to enucleate the contribution due to the correlations. The evaluation formula gets split  into three terms: the first  giving the zero correlations CVA value, i.e. the value for the independent case, the second and the third terms coming from the correlation of the intensity process respectively with the asset's price and the stochastic volatility. We call this the ``first order representation formula" since it involves the first order derivatives of the default free price.
This expression points out the contribution of the correlation, but it does not identify it explicitly. Were it needed, one can enlighten the role played by each process by applying again It\^ o's formula  to the default-free price, getting a so called ``second order representation formula" . The computations become more involved, but  this second expression captures more accurately the behavior of the correlations when these are parametrized.

More in detail, we model  the intensity  either by a Vasicek or a CIR  process each  coupled   with the SABR, Hull \& White and Heston stochastic volatility models. By consequence, the representation formula is rewritten explicitly in terms of the correlation coefficients among the drivers of the SDE's describing the model, in particular  between the asset's price and intensity ($\rho$) and  between the intensity and the volatility ($\nu$). The first and second order formulas we obtain  correspond to a first or second order expansion of the CVA with respect to the correlation parameters and this approach might be extended to any order at cost of a larger computational effort. These are approximated by ``ad hoc" techniques for each case and  the numerical accuracy and computational speed we achieve makes this technique a valid alternative to Monte Carlo methods.

To clarify the role of $\rho$, let us consider an investor $A$ buying from a bank $B$ a call option written on the asset of a bank $C$ competing with $B$ on a large market. If $B$ is going to default,  the intensity $\lam_t$ is going to have a large value and $\rho>0$ will reflect the propension of the asset $C$ to increase its value accordingly, due to gaining of market positions. Therefore $A$ will have in the portfolio a call contract which is deep in the money but with a counterparty proximal to default, describing a  wrong way risk (WWR) situation.

Finally, we remark  that our method can be  straightforwardly extended  to include a stochastic interest rate  and  its numerical accuracy and  speed makes it interesting  to apply it to more general XVA  evaluation in a future research.

In the next two sections, we first introduce the theoretical framework for CVA evaluation and then we provide our general representation formulas for a Markovian setting. In section 4 we specialize  the first order formula for the three stochstic volatility models associated with the two intensity models and we suggest an appropriate approximation technique for each of them. In section 5 we carry out the same discussion for the second order formula only for the SABR and Heston models coupled with a CIR intensity model. These are the most interesting cases and we decided to limit ourselves to those for the sake of exposition. The numerical discussion of our method is to be found in the last section.

\section{CVA Evaluation of Defaultable European Claims}
\label{sec1}

Let  $[0,T]$  be  a finite time interval and $(\Omega, \F, P)$ a complete probability space  endowed with a filtration  $\{\Ft\}_{t\in [0,T]}$, augmented with the
$P-$null sets and made right continuous. We also assume that all the processes have a c\'adl\'ag version.

The  market model will be described by the interest rate process $r_t$ and by a process  $X_t$  representing an asset log-price whose dynamics will be specified later.  The asset price may depend also on other stochastic factors.  We shall assume

\begin{itemize}
\item that the filtration $\{\Ft\}_{t\in [0,T]}$ is rich enough to support all the aforementioned  processes;

\item to be  in absence of arbitrage;

\item that the given probability  $P$ is a risk neutral measure, already selected by some criterion.

\end{itemize}

So we  denote by $S_t=\mathrm e^{X_t}$  the  asset price and by
$B(t,s)= \mathrm e^{-\int_t^s r_u du}$ the forward discounting factor.
In this market,  a defaultable European contingent claim paying $f(X_T)$ at maturity is traded, where $f$ is some function that will be specified later.
We denote by $\tau$ (not necessarily a stopping time w.r.t. the filtration $\Ft$) the default time  of the contingent claim and by   $Z_t$  an $\Ft-$measurable bounded recovery process.

To properly evaluate this type of derivative we need to include the information generated by the default time. We denote by $\Gt$ the progressively enlarged filtration, that makes $\tau$ a $\Gt -$stopping time, that is $\Gt=\Ft \lor \sigma(\{\tau\le t\})$. Hence, denoting by
$H_t = \mathbf 1_{\{\tau \le t\}}$ and by $
\Ht$ its natural filtration, we choose $\Gt=\Ft\lor\Ht$.

We make the fundamental assumption, known as the  H-hypothesis (see e.g. \cite{GJLR10} and \cite{G14} and the references therein), that

\medskip

\noindent
(H)\qquad \qquad\qquad \qquad  Every $\Ft-$martingale remains a $\Gt-$martingale.\hfill

\medskip \noindent
Under this assumption  $B(t,s)S_s, \, s\ge t$ remains  a $\Gs-$martingale under the unique extension of the risk neutral probability to the filtration $\Gs$.
(To keep notation light, we do not indicate  the probability we use for the  expectations, assuming to be  the  corresponding one to the filtration in use).

In this setting,  the price a defaultable claim, with final value  $f(X_T)$,  default time $\tau$ and recovery process $\{Z_t\}_t$ is given by
\begin{equation}
\label{eqvalue1}
c^d(t,T) = \E[B(t,T) f(X_T) 1_{\{\tau > T\}} + B(t,\tau) Z_{\tau} 1_{\{t < \tau \leq T\}} |\Gt],\quad t\in [0,T],
\end{equation}
while the corresponding default free  value  is
$
c(t,T) = \E[B(t,T)   f(X_T) |\Ft].
$

In many situations, investors do not know the default time and they may observe only whether it happened or not. The actual observable quantity is the asset price,  therefore  it  is interesting  to write the pricing formula \eqref{eqvalue1} in terms of $\Ft$, rather than in terms of $\Gt$. For that we have  the following key Lemma, see \cite{BJR} or \cite{BCB}.

\begin{lemma} For any integrable $\G-$measurable r.v. $Y$, the following equality holds
\begin{equation}
\label{key}
\E\Big[\mathbf 1_{\{\tau>
t\}}Y|\Gt\Big]=P(\tau>t|\Gt)\frac{\E\Big[\mathbf 1_{\{\tau>
t\}}Y|\Ft\Big]}{P(\tau>t|\Ft)}.
\end{equation}
\end{lemma}

Applying this lemma to the first and the second term of \eqref{eqvalue1}
and recalling that $1-H_t=\mathbf 1_{\{\tau>t\}}$ is $\Gt-$measurable, we obtain
\begin{eqnarray}
\label{key1}
\E[B(t,T) f(X_T) 1_{\{\tau > T\}}|\Gt]&=&\mathbf 1_{\{\tau>t\}}
\frac{\E[B(t,T) f(X_T) 1_{\{\tau > T\}}|\Ft]}{P(\tau>t|\Ft)}\\
\label{key1.2}
\E[B(t,\tau) Z_{\tau} 1_{\{t < \tau \leq T\}} |\Gt]&=&\mathbf 1_{\{\tau>t\}}\frac{\E[B(t,\tau) Z_{\tau} 1_{\{t < \tau \leq T\}} |\Ft]}{P(\tau>t|\Ft)},
\end{eqnarray}
which  may be made more explicit by following the hazard process  approach.

We denote  the conditional distribution of the default time  $\tau$ given $\Ft$ by
\begin{eqnarray}
F_t=P(\tau\leq t|\Ft), \qquad \forall \, t\ge 0,
\end{eqnarray}
whence, for $u\ge t$,  $P(\tau\leq u|\Ft)=\E(P(\tau\leq u|\Fu)|\Ft)=\E(F_u|\Ft)$. We also assume that $F_t(\omega)<1$ for all  $t>0$ to well define the so called hazard process
\begin{equation}
\label{risk}
\Gamma_t:=-\ln(1-F_t)\quad\Rightarrow \quad F_t= 1 - \mathrm e^{-\Gamma_t}\quad \forall \, t>0, \qquad  \Gamma_0=0.
\end{equation}

With this notation, we rewrite (\ref{key1})  as
$$
\begin{aligned}
&\E[B(t,T) f(X_T) 1_{\{\tau > T\}}|\Gt]=\mathbf 1_{\{\tau>t\}}
\E[B(t,T) f(X_T) 1_{\{\tau > T\}}|\Ft]\mathrm e^{\Gamma_t}\\
=&\mathbf 1_{\{\tau>t\}}\E[\E[B(t,T) f(X_T) 1_{\{\tau > T\}}|\FT]|\Ft]\mathrm e^{\Gamma_t}
=\mathbf 1_{\{\tau>t\}}
\E[B(t,T) f(X_T) \E[1_{\{\tau > T\}}|\FT]|\Ft]\mathrm e^{\Gamma_t}\\
=&
\mathbf 1_{\{\tau>t\}}\E[B(t,T) f(X_T) \mathrm e^{-\Gamma_T}\mathrm e^{\Gamma_t}|\Ft]=\mathbf 1_{\{\tau>t\}}\E[B(t,T) f(X_T) \mathrm e^{-(\Gamma_T-\Gamma_t)}|\Ft].
\end{aligned}
$$
Assuming that $B(t, \cdot)Z. $ is a bounded $\F-$ martingale (which is usually the case), we can
treat the expectation in \eqref{key1.2}, applying an extension of Proposition 5.1.1 of \cite{BR}, as developed in \cite{ARS1}.

\begin{proposition}\label{extension}
Let $M$ be a bounded,  $\cF$-martingale. Then

\begin{enumerate}

\item[(i)] for any $t \leq T$
\begin{equation}\label{eq1}
\E[M_{\tau} \mathbf{1}_{\{t< \tau\leq T\}}  |\cF_t] = \E\Big [\int_t^T M_{u-} dF_u + \sum_{t < u \leq T} \Delta M_u \Delta H_u |\cF_t\Big ],
\end{equation}
where by $\Delta X_s$ we denoted  $ X_s-X_{s-}$,  for any process $X$.
\item[(ii)] If moreover $M$ and $H$ do not have simultaneous jumps, then
\begin{equation}
\label{eq2}
\E[M_{\tau} \mathbf{1}_{\{t< \tau\leq T\}}  |\cF_t] = \E[\int_t^T M_{u -}dF_u |\cF_t].
\end{equation}

\item[(iii)] If either $M$ is  $\cF-$predictable  or $M$ and $H$ do not have simultaneous jumps and $F$ is predictable, then
\begin{equation}
\label{eq3}
\E[M_{\tau} \mathbf{1}_{\{t< \tau\leq T\}}  |\cF_t] = \E[\int_t^T M_u dF_u |\cF_t].
\end{equation}

\end{enumerate}

\end{proposition}

\begin{rem}
The processes $H$ and $B(t, \cdot)Z. $ do not have simultaneous jumps, since $B(t, \cdot)Z. $ is c\'adl\'ag  and $\tau$ is not measurable with respect to $\{\cF_t\}$, so it is not a $\cF-$stopping time. Jump times of    c\'adl\'ag process are stopping times (see \cite{Sok}), hence the two processes cannot jump simultaneously.
\end{rem}

Keeping the previous remark in mind and assuming that  the hazard process  is differentiable with derivative $\lambda_t$, called the intensity process,
$
\Gamma_t = \int_0^t \lambda_u du,
$  \eqref{key1.2} becomes
$$
\begin{aligned}
&\E[B(t,\tau) Z_{\tau} 1_{\{t < \tau \leq T\}} |\Gt]\!=\!\!\mathbf 1_{\{\tau>t\}} \E[B(t,\tau) Z_{\tau} 1_{\{t < \tau \leq T\}} |\Ft]\mathrm e^{\Gamma_t}\\
=&\!\!\mathbf 1_{\{\tau>t\}}\E\Big [\!\int_t^T\!\!\!\! B(t,s) Z_sdF_s |\Ft\Big ]\mathrm e^{\Gamma_t}=\mathbf 1_{\{\tau>t\}}\E\Big [\!\int_t^T B(t,s) Z_s\mathrm e^{-\Gamma_s}d\Gamma_s |\Ft \Big ]\mathrm e^{\Gamma_t}\\
=&\mathbf 1_{\{\tau>t\}}\E\Big [\int_t^T B(t,s) Z_s\mathrm e^{-(\Gamma_s-\Gamma_t)}d\Gamma_s |\Ft\Big ].
\end{aligned}
$$
since  $\Gamma$ is continuous and  it is not going to charge any jump. Hence the pricing formula \eqref{eqvalue1}  becomes
\begin{equation} \label{defaultprice}
\!c^d(t,T)\! = \!\mathbf 1_{\{\tau > t\}} \!\left[ \E\Big (\mathrm e^{-\!\int_t^T (r_s\!+\lambda_s) ds}f(X_T) \!+\!\!\int_t^T\!\!\! \!Z_s\lambda_s \mathrm e^{-\!\int_t^s (r_u\!+\lambda_u) du} ds| \Ft\Big ) \right]\! ,
\end{equation}
recovering  Lando's formulas (3.1) and (3.3) in \cite{Lando98}.

This formula can be specialized even further assuming  fractional recovery (\cite{DS}),  $Z_t = R c(t,T)$  for some $0\le R< 1$ and using the Optional Projection Theorem  (see \cite{N}  Theorem 4.16)
\begin{eqnarray*}
&&
\E[\int_t^T Z_{s} \lambda_s \mathrm e^{-\int_t^s (r_u+\lambda_u) du} ds| \Ft] = R   \E\Big [\int_t^T c(s,T) \lambda_s \mathrm e^{-\int_t^s (r_u+\lambda_u) du} ds| \Ft\Big ]\\
& =&
R  \E\Big [\int_t^T \E\Big (\mathrm e^{-\int_s^T r_u du} f(X_T)|\Fs\Big ) \lambda_s \mathrm e^{-\int_t^s (r_u+\lambda_u) du} ds| \Ft\Big  ] \\
&= & R\E\Big [\mathrm e^{-\int_t^T r_u du} f(X_T)  (1-\mathrm e^{-\int_t^T \lambda_u du})|\Ft\Big ].
\end{eqnarray*}

Putting the two pieces together, we finally obtain
\begin{equation}
\label{price2}
\begin{aligned}
c^d(t,T) = \mathbf 1_{\{\tau > t\}}\Big[R \E\Big(\mathrm e^{-\int_t^T r_udu} f(X_T) | \Ft \Big)
+ (1-R) \E\Big(\mathrm e^{-\int_t^T (r_u+\lambda_u) du} f(X_T) | \Ft \Big )\Big ],
\end{aligned}
\end{equation}
which was used also by Fard in \cite{Fard15} with $f(x)=(\mathrm e^x-K)^+$ and that can be interpreted as a convex combination of the default free price and the price with default.

As a consequence we have an expression also for the unilateral CVA, defined as difference between the default free price and the adjusted price
\begin{equation}
\label{cva0}
\begin{aligned}
CVA(t) :=& \mathbf 1_{\{\tau > t\}} [c(t,T) - c^d(t,T)]\\
=&\mathbf  1_{\{\tau > t\}} (1-R) \E\Big [\mathrm e^{-\int_t^T r_udu} f(X_T) (1-\mathrm e^{-\int_t^T \lambda_u du}) | \Ft\Big ].
\end{aligned}
\end{equation}
It is immediate to notice that if $\lam_t$ is independent from $(X_t,r_t)$, then
\begin{equation}
\label{ind2}
\E\Big [\mathrm e^{-\int_t^T (r_s + \lam_s) ds}f(X_T)|\Ft\Big ] = \E\Big[\mathrm e^{-\int_t^T r_s ds}f(X_T)|\Ft\Big ] \E\Big [\mathrm e^{-\int_t^T \lam_s ds}|\Ft\Big ].
\end{equation}
Of course, the computability of the expectations will depend heavily on the modeling choices one makes for $\lambda$.

\section{A representation formula for the CVA in stochastic volatility models }
\label{sec 2}

In this section we consider  a family of stochastic volatility models and, for the sake of simplicity,  from now on we assume zero fractional recovery ($R=0$) and risk free spot rate $r=0$. These are not restrictive assumptions, since the following discussion can be easily extended to the case of $0<R<1$ and $r$ a deterministic function of time. With the same techniques,  increasing the dimensionality of the problem also a stochastic interest rate might be considered.
Our market model is therefore given by the asset log-price  $X_t$ and  a stochastic factor, $Y_t$, determining the volatility process
\begin{eqnarray}
\label{SDEsystem0}
dX_t &=&-\frac{a^2(Y_t)}{2} \mathrm e ^{-2(1-\gamma)X_t}dt + a(Y_t) \mathrm e ^{-(1-\gamma)X_t}dB^1_t, \qquad 0<\gamma \le 1\\
\label{SDEsyst}
dY_t &=& b(t, Y_t) dt + c(t, Y_t) dB^2_t\\
\nonumber
B^1_t &= &\eta B_t + \sqrt{1-\eta^2} Z_t, \  \ \ B^2_t = B_t,
\end{eqnarray}
where $B_t$ and $Z_t$ are independent Brownian motions, $0\le\eta^2<1$,  $b, c: [0,T] \times \mathbb R^+\longrightarrow \mathbb R^+$ and $a: \mathbb R^+\longrightarrow \mathbb R^+$ are deterministic functions  so that  \eqref{SDEsystem0}-\eqref{SDEsyst} admits a unique strong solution.

We remark that, because of the deterministic coefficients, the pair $(X_t, Y_t)$ as well as the  stochastic factor $Y_t$ are  Markovian processes, which implies that the claim's price will be a deterministic function of these processes.
We remark that, with the right choice of functions, several popular models are included in the above formulation. As a matter of fact
\begin{eqnarray*}
\!\!\!\!\!\!\!&&\!\!\!\!\!\!\!\textrm {SABR model: } \gamma \in (0,1), \,\, a(y)=y,\,\,b(t,y)= 0, \,\,\,c(t,y)=cy,\,\, c>0\\
\!\!\!\!&&\!\!\!\!\!\!\!\textrm {Hull \& White model: }\gamma =1, \, a(y)=y,\,b(t,y)= b(t)y, \, c(t,y)=c(t)y,\, b,c:[0,T]\rightarrow \mathbb R^+\, \textrm{bounded}\\
 \!\!\!\!\!\!\!&&\!\!\!\!\!\!\!\textrm {Heston model: }\gamma =1, \,\,a(y)=\sqrt y,\,\, b(t,y)= k(\theta- y), \,\,c(t,y)=c\sqrt y, \quad k,\theta>0, c^2<2k\theta.
\end{eqnarray*}
From now on, we take the shorter notation $E_t$ to denote conditional expectation w.r.t.  the filtration generated by the processes under consideration
and let us consider the CVA problem given by formula \eqref{cva0},  for $R=0$. In other words we have to evaluate the risk-neutral expectation
\begin{equation}
\label{ev1}
\E_t\left[(1-N^t_T) U(T,X_T,Y_T)\right]\mathbf 1_{\{\tau > t\}},\quad \textrm{where} \quad N^t_s:=\E_s\Big( \e^{-\int_t^T\lambda_udu }\Big)
\end{equation}
 for $t\le s\le T$ and by $U(s,X_s,Y_s)$ we denoted the default free price of the European claim whose payoff is $U(T,X_T,Y_T)=f(X_T)$. The default free price $U$ is,  by no arbitrage, a martingale  and consequently
applying It\^{o}'s formula we have
\begin{equation}
\begin{aligned}
dU(s,X_s,Y_s))&=\cL U(s,X_s, Y_s)ds\\
&+ a(Y_s) \mathrm e ^{-(1-\gamma)X_s}\partial_x U(s,X_s, Y_s) dB^1_s + c(s, Y_s)\partial_{y} U(s,X_s, Y_s)dB^2_s\\
&=\partial_x U(s,X_s, Y_s) dM^X_s +  \partial_yU(s,X_s, Y_s)dM^Y_s,
\end{aligned}
\end{equation}
where  $M^X$ and $M^Y$ are the respective martingale parts of $X$ and $Y$ and
$$
\mathcal{L}=\partial_{s}+ \frac{a^2(y) \mathrm e ^{-2(1-\gamma)x}}2( \partial^2_{xx}-\partial_x)
+ \frac{c^2(s, y)}2\partial^2_{yy}\\
+\eta a(y) c(s, y)\mathrm e ^{-(1-\gamma)x}\partial^2_{xy}
+ b(s,y) \partial_{y}
$$
and
 $U$ verifies the PDE
\begin{equation}
\label{pde}
\begin{cases}&
\cL U(s,x,y)=0\\
&U(T,x,y)= f(x).
\end{cases}
\end{equation}
By integration by parts and making use of (\ref{pde}) we get the following  basic representation formula of the difference between the `classical price' and the adjusted  price
on the event $\{ \tau>t\}$
\begin{equation}
\label{cva1bis}\boxed{
\begin{aligned}
CVA(t)=&\E_t\left[(1-N^t_T) U(T,X_T,Y_T)\right]=(1\!-\!N^t_t )U(t,x,y)\\
-&\E_t\Big [\!\int_t^T\!\!\! \!\partial_x U(s,X_s,Y_s)d\langle N^t\! ,X\rangle_s\Big ]
\!- \E_t\Big [\!\int_t^T\!\!\! \partial_{y} U(s,X_s,Y_s)d\langle N^t\!,Y\rangle_s\Big],
\end{aligned}}
\end{equation}
that is the uncorrelated term plus two extra terms that come from the correlation between the asset and the default and the volatility and the default.
\begin{rem}
We may view  evaluation \eqref{ev1} from a different perspective:
$$
\textrm{CVA}(t)=
\E_t\left[(1-N^t_T) u(T,X_T,v^2_T)\right]\mathbf 1_{\{\tau > t\}},
$$
where $ v^2_t:=V^2_t(T-t)$,
$\sigma_t=a(Y_t)$ is the stochastic volatility process and $V_t^2$  denotes the adapted averaged variance process (or zero-strike variance swap)
$$
V_t^2=\frac{1}{T-t}\E_t\Big (\int_t^T\sigma^2 ds\Big )=\frac{1}{T-t}\E_t\Big (\int_t^Ta^2(Y_s) ds\Big ).
$$
The process $v^2$ is called the variance swap and it more easily estimated from the market data rather than non tradeable stochastic factor $Y$. These two proceses are strictly connected, and we assume
$$
\textrm{ (H1) }\quad
\textrm{there exists an invertible function } d(t,y)\in C^{1,2}([0,T] \times \mathbb R^+) \textrm{  so that } v^2_t= d(t,Y_t),\qquad \qquad\qquad\qquad
$$
and we may equivalently write the evaluations by means of $v^2$.
Indeed,  by (H1) $(X, v^2) $ is still a Markovian pair and  the function $u$ verifies an equivalent partial differential equation
\begin{equation}
\label{pde1}
\Big [\partial_{s}+f(s,z)[ \frac{ \mathrm e ^{-2(1-\gamma)x}}2(\partial^2_{xx}-\partial_x)-\partial_z]+\eta \mathrm e ^{-(1-\gamma)x} g(s,z)\partial_{xz}^2+h(s,z)\partial_{zz}^2\Big ]u(s, x, z)=0
\end{equation}
with
\begin{equation*}
\begin{aligned}
f(s,v_z)&=a^2(d^{-1}(s,z)), \quad
g(s,z)= a^2(d^{-1}(s,z))\partial_y d(s,d^{-1}(s,z)), \\
h(s,z)&= \frac{c^2(s, d^{-1}(s,z))}2(\partial_y d(s, d^{-1}(s,z)))^2,
\end{aligned}
\end{equation*}
arriving at
\begin{equation}
\label{cva1}
\begin{aligned}
&CVA(t)=\E_t\left[(1-N^t_T) u(T,X_T,v_T^2)\right]\\
=&(1\!-\!N^t_t )u(t,X_t,v_t^2)-\E_t\Big [\!\int_t^T\!\!\! \!\partial_x u(s,X_s,v_s^2)d\langle N^t\! ,X\rangle_s\Big ]
\!- \E_t\Big [\!\int_t^T\!\!\! \!\partial_{z} u(s,X_s,v_s^2)d\langle N^t\!,v^2\rangle_s\Big].
\end{aligned}
\end{equation}
\end{rem}
Hypotheses (H1) is not restrictive, as a matter of fact it is verified by all the models we are interested in
\begin{eqnarray*}
&&\textrm{SABR}\quad
\begin{aligned}\label{SABR1}
&v^2_t = d(t, Y_t)=Y^2_t\phi(t,T), \quad \phi(t,T):= \frac{\mathrm e^{c^2(T-t)} - 1}{c^2}, \quad c(t, Y_t)=c Y_t \\
\Rightarrow \quad & \partial_y d (t, Y_t)=2 Y_t\phi(t,T),\quad Y_t=\frac {v_t}{\sqrt{ \phi(t,T)}},
\end{aligned}\\
&&\textrm{H  \& W}\quad
\begin{aligned}\label{HW1}
&v^2_t =  Y^2_t  \phi(t,T),
 \quad  \phi(t,T):= \int_t^T\mathrm e^{\int_t^s [b(u)+ c(u)]du}ds, \quad c(y, Y_t)= c(t) Y_t\\
 \Rightarrow \quad &\partial_y d (t, Y_t)=2 Y_t\phi(t,T), \quad Y_t=\frac {v_t}{\sqrt{ \phi(t,T)}},
\end{aligned}\\
&&\textrm{Heston}\quad
\begin{aligned}
&v^2_t = \theta (T-t) +(Y_t-\theta)\phi(t,T)\quad \phi(t,T):=\frac{1-\mathrm e^{-k(T-t)}}{k},\quad c(t,Y_t)=c\sqrt{Y_t}\\
 \Rightarrow\quad &\partial_y d (t, Y_t)=\phi(t,T),  \quad Y_t=\frac{v_t^2-\theta(T-t)}{\phi(t,T)}+\theta.
\end{aligned}
\end{eqnarray*}

\begin{rem}
Since  $\sigma_t = a(Y_t)$, when $a$ is also an invertible and differentiable function, we have that  $u(t,X_t,v_t^2)=   U(t,X_t, Y_t) = U(t,X_t, a^{-1}(\sigma_t) )$, consequently
$$
Vega(t) = \partial_\sigma U((t,X_t, a^{-1}(\sigma_t) )= \partial_z u(t,X_t,v_t^2)\partial_y d(t, Y_t)\partial_\sigma a^{-1}(\sigma_t),
$$
whence we may give an approximation of the above expression \eqref{cva1}, by freezing the integrands at the inital time $t$
\begin{equation}
\label{approxim}
CVA(t)\!
\approx\!(1\!-\!N^t_t )u(t,X_t,v_t^2)
-Delta(t)  \E_t\Big [\langle N^t,X\rangle\Big |_t^T\Big]\!
-\frac{Vega(t)}{\partial_yd(t, Y_t)\partial_\sigma a^{-1}(\sigma_t)}  \E_t\Big [\langle N^t\!,v^2\rangle\Big |_t^T\Big ]
\end{equation}
since
\begin{eqnarray}
\partial_x u(t,X_t,v_t^2)&=&\partial_x U(t,X_t,Y_t)=Delta(t) \\
 \partial_z u(t,X_t,v_t^2)&=& \frac {Vega(t)}{\partial_yd(t, d^{-1}(t,v^2_t))\partial_\sigma (a^{-1})(\sigma_t)}.
\end{eqnarray}
\end{rem}
When specializing to the three models we are interested in,  assuming some specific dynamics for $\lambda$, the quantities  appearing in \eqref{cva1bis}  or  \eqref{cva1} might be explicitly computed in terms of the correlation parameters between the intensity and the other two processes, which is what we are going to do in the next sections.

The representation formula \eqref{cva1bis} is quite crucial, because it allows to exploit an explicit expression of the martingale $N^t$, whenever possible.
This happens, for instance when considering affine models for the dynamics of $\lambda$, in this case  we may resort to the bond pricing theory and  we may obtain that
\begin{equation}
\label{bond0}
N^t_s= \E_s\Big ( \mathrm e^{-\int_t^T\lambda_udu } \Big)=\mathrm e^{-\int_t^s\lambda_udu }E_s\Big ( \mathrm e^{-\int_s^T\lambda_udu } \Big)= \mathrm e^{-\int_t^s\lambda_udu }\e^{\varphi(T-s)\lam_s+ \psi(T-s)}
\end{equation}
for some deterministic differentiable  functions $\varphi$ and $\psi$ of time to maturity, which implies that
$$
\begin{aligned}
dN^t_s=& -N^t_s\Big \{ [ \lam_s +\varphi'(T-s)\lam_s + \psi'(T-s) ]ds +\varphi(T-s)d \lam_s-\frac 12\varphi^2(T-s)d\langle\lam,\lam\rangle_s \Big \}\\
=&- N^t_s  \varphi(T-s)dM^\lam_s\\
d\langle X, N^t\rangle_s=&-N^t_s  \varphi(T-s)d\langle X, \lam\rangle_s\\
d\langle Y, N^t\rangle_s=&-N^t_s  \varphi(T-s)d\langle Y, \lam\rangle_s,
\end{aligned}
$$
where $M^\lam$ denotes the martingale part of the process $\lam$. Representation \eqref{cva1bis} hence becomes
\begin{equation}
\label{cva1.2}
\begin{aligned}
&CVA(t)=\E_t\left[(1-N^t_T) U(T,X_T,Y_T)\right]=(1\!-\!N^t_t )U(t,x,y)\\
+&\E_t\Big [\!\int_t^T\!\!\! \!\partial_x U(s,X_s,Y_s)N^t_s  \varphi(T-s)d\langle X, \lam\rangle_s\Big ]
\!+ \E_t\Big [\!\int_t^T\!\!\! \partial_yU(s,X_s,Y_s)N^t_s  \varphi(T-s)d\langle Y, \lam\rangle_s\Big].
\end{aligned}
\end{equation}
The quadratic covariations will express the correlation between the processes of the model. We remark, though, that correlations are still present between
$ \partial_x U(s,X_s,Y_s)$, $ \partial_y U(s,X_s,Y_s)$ and the processes defined by
$$
H^x_s= N^t_s  \frac{d\langle X, \lam\rangle_s}{ds}, \qquad H^y_s=N^t_s \frac{d\langle Y, \lam\rangle_s}{ds},
$$
so one might think of refining the above formula, by applying It\^o's formula and integration by parts once again.

By differentiating \eqref{pde} with respect to $x$ and $y$, we may conclude that $\partial_x U$ and $\partial U_y$ verify  the following
system of  PDE's
\begin{equation}
\label {pde-2}
\begin{cases}
&\cL (\partial _xU )-(1-\gamma) \cL^{1,x} (\partial _xU)=0, \\
 & \cL( \partial _yU) + \cL^{1,y} (\partial _yU)+ aa'(y) \mathrm e^{-2(1-\gamma)x}(\partial_x-1)(\partial_x U)=0,
\end{cases}
\end{equation}
where
$$
\begin{aligned}
\cL^{1,x} &= a^2(y) \mathrm e^{-2(1-\gamma)x}(\partial_x -1) +\eta c(s,y) a(y)  \e^{-(1-\gamma)x}\partial_y \\
\cL^{1,y} &= cc'(s,y)\partial_y + \eta[ a'(y) c(s,y)+ a(y) c'(s,y)]  \e^{-(1-\gamma)x}\partial_x+ b'(s,y)
\end{aligned}
$$
Consequently,  applying It\^o calculus  we have
$$
\begin{aligned}
&\E_t\Big [\int_t^T\!\!\! \varphi(T\!-s) \partial_x U(s,X_s,Y_s)H^x_sds\Big ]= \partial_x U(t,x,y)H^x_t \int_t^T\!\!\! \varphi(T\!-s)ds\\
&+ \int_t^T\!\!\! \varphi(T\!-s) \E_t\Big [\!\int_t^s\!\!\!\!\! H^x_u d(\partial_x U(u,X_u,Y_u))
+\!\!\!\int_t^s\!\!\!\! \partial_x U(u,X_u,Y_u) dH^x_u + \int_t^s d\langle H^x,   \partial_x U\rangle_u \Big ]ds\\
=& \partial_x U(t,x,y)H^x_t \int_t^T\!\!\! \varphi(T\!-s)ds+ \int_t^T\!\!\! \varphi(T\!-s)\!\int_t^s\!\!\!\!\!  \E_t\Big [H^x_u \cL (\partial _xU )(u,X_u,Y_u)\Big ]duds\\
&+\int_t^T\!\!\! \varphi(T-s)\E_t\Big [\int_t^s\!\!\!\! \partial_x U(u,X_u,Y_u) dH^x_u\Big ] ds\\
&+\int_t^T\!\!\! \varphi(T-s)\E_t\Big [\int_t^s\!\!\!\partial_{xx}U (u, X_u, Y_u) d\langle H^x, X\rangle_u+\int_t^s\!\!\!\partial_{xy}U (u, X_u, Y_u) d\langle H^x, Y\rangle_u\Big ] ds\\
=& \partial_x U(t,x,y)H^x_t \int_t^T\!\!\! \varphi(T\!-s)ds+(1-\gamma)\int_t^T\!\!\! \varphi(T\!-s)\!\int_t^s\!\!\!\!\!  \E_t\Big [H^x_u   \cL^{1,x}  (\partial _xU )(u,X_u,Y_u)\Big ]duds\\
&+\int_t^T\!\!\! \varphi(T-s)\E_t\Big [\int_t^s\!\!\!\! \partial_x U(u,X_u,Y_u) dH^x_u\Big ] ds\\
&+\int_t^T\!\!\! \varphi(T-s)\E_t\Big [\int_t^s\!\!\!\partial_{xx}U (u, X_u, Y_u) d\langle H^x, X\rangle_u+\int_t^s\!\!\!\partial_{xy}U (u, X_u, Y_u) d\langle H^x, Y\rangle_u\Big ] ds
\end{aligned}
$$
Similarly for the other  term we obtain
$$
\begin{aligned}
 &\E_t\Big [\!\int_t^T\!\!\! \partial_yU(s,X_s,Y_s)N^t_s  \varphi(T-s)d\langle Y, \lam\rangle_s\Big]= \partial_y U(t,x,y)H^y_t \int_t^T\!\!\! \varphi(T\!-s)ds\\
&+\int_t^T\!\!\! \varphi(T\!-s)\!\int_t^s\!\!\!\!\!  \E_t\Big [H^y_u (   \cL^{1,y} (\partial _yU)+ aa'(Y_u) \mathrm e^{-2(1-\gamma)X_u}(\partial_x-1)(\partial_x U)(u,X_u,Y_u))\Big ]duds\\
&+\int_t^T\!\!\! \varphi(T-s)\E_t\Big [\int_t^s\!\!\!\! \partial_y U(u,X_u,Y_u) dH^y_u\Big ] ds\\
&+\int_t^T\!\!\! \varphi(T-s)\E_t\Big [\int_t^s\!\!\!\partial_{xy}U (u, X_u, Y_u) d\langle H^y, X\rangle_u+\int_t^s\!\!\!\partial_{yy}U (u, X_u, Y_u) d\langle H^y, Y\rangle_u\Big ] ds.
\end{aligned}
$$
Plugging these expressions into  \eqref{cva1.2}, we obtain a second order representation formula for CVA
\begin{equation}
\label{cva1.3}
\boxed{
\begin{aligned}
&\textrm{CVA}(t)=(1\!-\!N^t_t )U(t,x,y)+  \Big [\partial_x U(t,x,y)H^x_t +  \partial_y U(t,x,y)H^y_t\Big ] \int_t^T\!\!\! \varphi(T\!-s)ds \\
&+(1-\gamma)\int_t^T\!\!\! \varphi(T\!-s)\!\int_t^s\!\!\!   \E_t\Big [H^x_u   \cL^{1,x}  (\partial _xU )(u,X_u,Y_u)\Big ]duds\\
&+\int_t^T\!\!\! \varphi(T\!-s)\!\int_t^s\!\!\!\!  \E_t\Big [H^y_u \Big (   \cL^{1,y} (\partial _yU)+ aa'(Y_u) \mathrm e^{-2(1-\gamma)X_u}(\partial_x-1)(\partial_x U)(u,X_u,Y_u)\Big )\Big ]duds\\
&+\int_t^T\!\!\! \varphi(T\!-s)\E_t\Big [\int_t^s\!\!\!\! \partial_x U(u,X_u,Y_u) dH^x_u\Big ] ds+\int_t^T\!\!\! \varphi(T-s)\E_t\Big [\int_t^s\!\!\!\! \partial_y U(u,X_u,Y_u) dH^y_u\Big ] ds\\
&+\int_t^T\!\!\! \varphi(T\!-s)\E_t\Big [\int_t^s\!\!\!\partial_{xx}U (u, X_u, Y_u) d\langle H^x, X\rangle_u+\int_t^s\!\!\!\partial_{xy}U (u, X_u, Y_u) d\langle H^x, Y\rangle_u\Big ] ds\\
&+\int_t^T\!\!\! \varphi(T\!-s)\E_t\Big [\int_t^s\!\!\!\partial_{xy}U (u, X_u, Y_u) d\langle H^y, X\rangle_u+\int_t^s\!\!\!\partial_{yy}U (u, X_u, Y_u) d\langle H^y, Y\rangle_u\Big ] ds.
\end{aligned}}
\end{equation}
In the next sections we will consider either the Vasicek or the CIR model for the intensity process and, after specializing  the above  representations in those cases for a call option, we are going to suggest appropriate approximation formulas, that are going to be discussed numerically in the last section.

 We remark that, even though the Vasicek model has already been used in the literature for the intensity process (see \cite{Fard15}), it has the drawback that it can become negative  with positive probability, making the process $\Gamma$ a bad candidate to represent a probability. This means that  this model can be employed only if the estimated parameter values imply that the probability $\lam$ may become negative is negligible.

\section{The first order approximation}

\label{sec3}

In this section we analyze the CVA   first order representation formula   \eqref{cva1bis} for a defaultable call option  with strike price given by $K= \mathrm e^\kappa, \, \kappa \in \mathbb R$,  for our three models when either a Vasicek  or a CIR dynamics is chosen for the intensity.

\subsection{The Vasicek intensity model}

We assume that the intensity process verifies
\begin{equation}
d\lam_s=q(\mu-\lam_s) ds + \sigma dB_s^3,
\end{equation}
with constants $q, \mu, \sigma>0$  and
$$
B^1_s= \eta B_s+ \sqrt{1-\eta^2} Z_s, \ \ \  B^2_s = B_s,\ \ \ \ B^3_s =\nu B_s+ \frac{\rho-\eta\nu}{\sqrt{1-\eta^2}} Z_s+ \sqrt{\frac{1-(\eta^2+ \nu^2+\rho^2)+2\nu\eta\rho}{1-\eta^2}} U_s,
$$
where ${W}_s=(Z_s, B_s, U_s)$ are independent Brownian motions, $(\nu, \eta,\rho)$ are such that
$$
\nu^2 <1, \,\,\, \eta^2< 1,\,\,\rho^2<1, \quad  \nu^2 +\rho^2+\eta^2<1+2\nu\eta\rho.,
$$
In this way we set up the correlations to be
$$
\langle B^1, B^2\rangle_s=\eta s, \quad \langle B^1, B^3\rangle_s=\rho s, \quad\langle B^2, B^3\rangle_s=\nu s.
$$
and to simplify notation  we set
$$\alpha =\frac{\rho-\eta\nu}{\sqrt{1-\eta^2}} \quad \textrm{and } \quad \beta=  \sqrt{\frac{1-(\eta^2+ \nu^2+\rho^2)+2\nu\eta\rho}{1-\eta^2}}  \quad \Rightarrow \quad \nu^2+\alpha^2+ \beta^2=1.
$$
Consequently,  for $t\le s \le T$, we have
\begin{equation}
\label{bondvas}
\begin{aligned}
N^t_s&=\mathrm e^{-\int_t^s\lambda_udu }\e^{\varphi(T-s)\lam_s+ \psi(T-s)}\qquad \textrm{with }\\
\varphi(T\!-s)&=-\frac{1-\mathrm e^{-q(T\!-s)}}q,\quad  \psi(T\!-s)=-[\mu-\frac{\sigma^2}{2q^2}](\varphi(T\!-s)+(T-s))-\frac{\sigma^2 \varphi^2(T\!-s)}{4q}
\end{aligned}
\end{equation}
and $N^t_t$  is the price of a zero coupon bond with spot rate $\lambda$.

\subsubsection{SABR - V}
\label{subsec3.1}

Chosen $0<\gamma < 1$, we have for $s>t$
\begin{equation}
\label{SABR2}
\begin{aligned}
dX_s &=-\frac{Y^2_s}{2} \mathrm e ^{-2(1-\gamma)X_s}ds + Y_s \mathrm e ^{-(1-\gamma)X_s}dB^1_s, \qquad X_t =x\\
dY_s&=cY_sdB^2_s, \qquad Y_t =y.
\end{aligned}
\end{equation}
Therefore
$$
\begin{aligned}
d\langle X, N^t\rangle_s=& -\rho\sigma  N^t_s \varphi(T-s)Y_s \mathrm e ^{-(1-\gamma)X_s}ds\\
d\langle Y, N^t\rangle_s=&-c \nu\sigma  N^t_s \varphi(T-s)Y_s ds,
\end{aligned}
$$
so   that  \eqref{cva1bis} becomes
\begin{equation}
\label{cva2}
\begin{aligned}
CVA(t)&=(1\!-\!N^t_t )U(t,x,y)+\rho\sigma \E_t\Big [\!\int_t^T\!\!\! \partial_x U(s,X_s,Y_s)N^t_s  \varphi(T-s)Y_s \mathrm e ^{-(1-\gamma)X_s}ds\Big ]\\
&+
c \nu\sigma \E_t\Big [\!\int_t^T\!\!\! \partial_y U(s,X_s,Y_s)N^t_s  \varphi(T-s)Y_sds\Big],
\end{aligned}
\end{equation}
which might be easily approximated by  freezing at the initial time all the terms containing $X$
\begin{equation}
\label{apprSV}
\begin{aligned}
CVA(t)&
\approx(1-N^t_t )U(t,x,y)
+ \rho\sigma\partial_x U(t,X_t,Y_t) \e ^{-(1-\gamma)x}  \int_t^T \varphi(T-s) E_t(N^t_s Y_s ) ds\\
&+ c\nu\sigma\partial_y U(t,x,y)  \int_t^T\varphi(T-s)E_t(N^t_s Y_s)ds.
\end{aligned}
\end{equation}
We remark that this choice forces the approximation to stay linear in $\rho$ and we will see later that this is well supported by the numerical simulations. Were it not the case, one might think of refining the calculations by building an approximation for the second order representation formula \eqref{cva1.3}.

Therefore the problem is reduced to having an expression for the default free evaluation $U$.
The power of the  SABR model lies in the fact that an explicit formula for the call price can be derived by substituting in the Black's formula the   SABR implied volatility, $\sigma_B(t,x,y, \kappa)$ (see \cite{Ha}) and in our case ($r=0$) we obtain
\begin{equation}
\label{black}
U(t, x, y)=\mathrm e^x\cN(d_1) - \mathrm e^\kappa \cN(d_2), \qquad d_{1,2}= \frac { x-\kappa \pm \frac{\sigma_B^2}2 (T-t)}{\sigma_B\sqrt {T-t}}.
\end{equation}
The  implied volatility may be efficiently approximated by truncating the formula (2.17a)    in \cite{Ha} up to the second order. In our setting this approximation  is
\begin{equation}
\label{sabr3}
\begin{aligned}
\tsigma_B(t, x, y):=&\frac y{ \mathrm e^{ (x+\kappa)\frac {1- \gamma}2 }  [1+ \frac{(1-\gamma)^2}{24}(x-\kappa)^2+\frac{(1-\gamma)^4}{1920}(x-\kappa)^4 ]}
\Big (\frac m {\delta(m)}\Big )\\
&\cdot\left \{1 +\Big [ \frac {(1-\gamma)^2}{24} \frac {y^2}{ \mathrm e^{ (x+\kappa)(1- \gamma) }}+\frac 14\frac{\eta\gamma  c y}{ \mathrm e^{ (x+\kappa)\frac {1- \gamma}2 } }+ \frac { 2- 3 \eta^2}{24}c^2\Big ] (T-t)\right \},
\end{aligned}
\end{equation}
where
\begin{equation}
m(x,y)=\frac c y  \mathrm e^{ (x+\kappa)\frac {1- \gamma}2}(x-k),\quad
\delta(m)=\ln \left(\frac{m-\eta+\sqrt{(m-\eta)^2+ 1-\eta^2}}{1-\eta}\right).
\end{equation}
We notice that actually $U(t,x,y)=U(t, x,\sigma_B(t,x,y))$, whence we may approximate $U$ by
$$
\tilde U(t,x,z)= U\Big (t,x,\tsigma_B(t,x,y)\Big)
$$
and we obtain
$$
\begin{aligned}
\partial_x\tilde U(t,x,y)=& (\partial_x U+ \partial_\sigma U\partial_x\tsigma_B) (t,x,
y)
=\mathrm e^x\Big [ \cN(d_1)+\sqrt{T-t} \cN'(d_1)\partial_x\tsigma_B\Big ](t,x,y)\\
\partial_y\tilde U(t,x,y)=&(\partial_\sigma U\partial_y \tsigma_B)(t,x, y)
=\sqrt{T-t}\,\mathrm e^x [\cN'(d_1)\partial_y\tsigma_B](t,x,y).
\end{aligned}
$$
We omit here the lengthy calculations of the two derivatives $\partial_x\tsigma_B$ and $ \partial_y \tsigma_B$, which we performed with the help of Matlab.
Inserting the above expressions in \eqref{apprSV}, we obtain
\begin{equation}
\label{sabrfin}
\begin{aligned}
&CVA(t)
\approx(1-\e^{\varphi(T-t)\lam+ \psi(T-t) } ) [\mathrm e^x\cN(d_1) - \mathrm e^\kappa \cN(d_2)](t, x, \kappa, \tsigma_B)\\
&+\rho \sigma\mathrm e^{\gamma x}\Big [ \cN(d_1)+\sqrt{T-t}\, \cN'(d_1)\partial_x\tsigma_B\Big ] \int_t^T \varphi(T-s)\E_t (N^t_sY_s)ds\\
&+ c\nu\sigma \sqrt{T-t}\,\e^x [\cN'(d_1)\partial_y\tsigma_B]
\int_t^T\varphi(T-s) \E_t(  N^t_sY_s) ds.
\end{aligned}
\end{equation}
Arrived at this stage, we might obtain a very handy approximation by simply freezing also all  the other processes at the initial time. As well, we might think of assuming $\nu=0$ (often done by practitioners) so that $ \E_t(  N^t_sY_s)=\E_t(  N^t_s)\E_t( Y_s)=y N_t^t $. Both  choices are rather crude, so we decided to push computations a little further to get a more accurate evaluation of this expectation.
Recalling the explicit expressions of $Y_s$ and $\lam_s$
 \begin{eqnarray*}
Y_s&=&y \mathrm e^{c(B^2_s-B^2_t) - \frac {c^2}2 (s-t)}\\
\lambda_s& =& \mu + (\lambda - \mu) \mathrm e^{-q(s-t)}+ \sigma\int_t^s  \mathrm e^{-q(s-u)}dB^3_u\\
\int_t^s\lam_u du &=&\mu(s-t) +(\lam - \mu) \frac {1- \mathrm e^{-q(s-t)}}q + \frac \sigma q \int_t^s (1- \mathrm e^{-q(s-u)}) dB^3_u,
\end{eqnarray*}
the definition of $\psi(T-s) $ and $\varphi(T-s)$ and that  $B^2_u= B_s$ and $B^3_u= \nu B_u+ \alpha Z_u+\beta U_u$ and  gathering the alike terms together, after some calculations we
arrive at the following expression
\begin{eqnarray*}
\E_t(  N^t_sY_s)
&=&y\mathrm e^{f_1(t,s)}\E_t\Big (\mathrm e^{\int_t^s (\sigma\nu \varphi(T-u)+ c)dB_u+ \sigma \int_t^s \varphi(T-u)(\alpha dZ_u+ \beta dU_u)}\Big)=y\mathrm e^{f_1(t,s)+\frac 12 f_2^2(t,s) },
\end{eqnarray*}
where we used  that the independence of the Brownian motions $B,Z,U$ implies that the exponent is Gaussian with density $\cN(0,  f_2^2(t,s) )$, with
$$
\begin{aligned}
f_1(t,s)&=-\mu(T-t)- \frac {c^2}2(s-t) + \frac{\sigma^2}{2q^2}(T-s)+ (\lam-\mu)\varphi(T-t)-\frac{\sigma^2}{4q^2}\frac{3+ \mathrm e^{-2q(T-s)}-4\mathrm e^{-q(T-s)}}q\\
f_2^2(t,s)&=[\frac{\sigma^2}{q^2} \! - 2\frac{ c\nu\sigma}q](s-t)- 2[\frac{ \sigma^2} {q^2}- \frac{ \sigma c \nu}q]\frac {\mathrm e^{-q(T-s)}- \mathrm e^{-q(T-t)}}q + \frac{\sigma^2}{q^2}\frac {\mathrm e^{-2q(T-s)}\!- \mathrm e^{-2q(T-t)}}{2q}\\
&+ c^2(s-t) .
\end{aligned}
$$

\subsubsection{Hull \& White - V}

\label{subsec3.2}

Here we have
\begin{equation}
\label{SDEsystemHW}
\begin{aligned}
dX_s&=-\frac{Y^2_s}{2} ds+Y_s dB^1_s\\
dY_s &= b(s)Y_s ds + c(s)Y_s dB^2_s
\end{aligned}
\end{equation}
and we remark that we assume $\eta<0$, since  the martingale nature of the price under a risk
neutral probability is guaranteed if and only if the correlation coefficient is non
positive, see \cite{HW}.
Since the martingale $N^t_s$ is defined by \eqref{bondvas},  for this model we have
\begin{eqnarray*}
d\langle X, N^t\rangle_s&=&- \rho\sigma  \varphi(T-s) N^t_s Y_s ds\\
d\langle Y, N^t\rangle_s&=&-\nu\sigma c(s)   \varphi(T-s)N^t_s Y_s ds,
\end{eqnarray*}
with   \eqref{cva1bis}
that might be approximated by
$$
\begin{aligned}
CVA(t)
\approx&(1-N^t_t )U(t,x,y)
+ \rho\sigma\partial_x U(t,x,y) \int_t^T\!\!\!  \varphi(T\!\!-\!s)E_t(N^t_s Y_s ) ds\\
&+\nu\sigma\partial_y U(t,x,y)  \int_t^T\!\!\!c(s)  \varphi(T\!\!-\!s)E_t(N^t_s Y_s)ds.
\end{aligned}
$$
As before, the problem is reduced to evaluating $E_t(N^t_s Y_s )$,  easily done since both processes are lognormally distributed
\begin{equation}
\label{finalHW1}
\begin{aligned}
&E_t(N^t_s Y_s )=
y\e^{ \Big [\frac{\sigma^2}{2q^2} \!- \!\mu\Big ](T\!\!-\!t)+ \Big [\lam\! -\!\mu +\frac{\sigma^2}{4q^2}
\Big (3\!- \mathrm e^{-q(T\!-\!t)}\Big )\Big ]\varphi(T\!\!-\!t)+ \int_t^s \! b(u) + \nu c(u)\varphi(T\!\!-\!u)]du }.
\end{aligned}
\end{equation}
Instead, it does not exist an explicit pricing formula for the call price in a Hull \& White stochastic volatility model, hence  to approximate those
$U(t,X_t,Y_t)$ and its derivatives
 we   employ the power series expansion introduced in \cite{AS}, leading to the following approximation
\begin{equation}
\label{appHW}
\begin{aligned}
U(t,x,y)&\approx \bar g_0(t,x,y)+\eta \bar g_1(t,x,y)\\
 \bar g_0(t,x,y)&=c_{BS}\Big (t, x,
\sqrt{\int_t^T\E(Y^2_s) ds}\Big )\\
\bar g_1(t,x,y)&= -\frac 1 y\mathrm e^{\kappa} \frac{\bar
d_2\cN'(\bar d_2)}{\E(\int_t^TY^2_s ds)} \int_t^T\!\int_s^Tc(s)\E\Big (Y^2_s
Y^2_u \Big)
du ds ,
\end{aligned}
\end{equation}
where
$$
\begin{aligned}
\E(Y^2_s) &= y^2 \mathrm e^{\int_t^s[ 2b(w)+ c^2(w)] dw}=:y^2\Gamma(t, s)
,\\
\E\Big (Y_s^2Y_u^2\Big )&= y^4 \mathrm e^{\int_t^s[4b(w)+6c^2(w)]dw+\int_s^u[2b(w)+c^2(w)]dw}
\\
\bar d_2 &=   \frac {(x-\kappa)- \frac {y^2}2\int_t^T\Gamma(t, s) ds}{y\sqrt{\int_t^T \Gamma(t, s)ds}}.
\end{aligned}
$$

\subsubsection{Heston - V}

\label{subsec3.3}

Here the market model is given by
\begin{eqnarray*}
\label{SDEsystemH}
dX_s &=&-\frac{Y_s}2ds + \sqrt{Y_s}dB^1_s\\
dY_s &=& k(\theta -Y_s) ds + c\sqrt{Y_s}dB^2_s
\end{eqnarray*}
and from  \eqref{bondvas}, for this model we have
$$
d\langle X, N^t\rangle_s=- \rho\sigma  N^t_s\varphi(T\!\!-\!s)\sqrt {Y_s }ds\qquad
d\langle Y, N^t\rangle_s=- \nu \sigma  c N^t_s \varphi(T\!\!-\!s)\sqrt{Y_s} ds,
$$
so that \eqref{cva1bis} might be approximated by
$$
CVA(t)
\approx (1-N^t_t )U(t,X_t,Y_t)
+\sigma\Big [\rho \partial_x U(t,X_t,Y_t)+ c \nu \partial_y U(t,X_t,Y_t) \Big] \int_t^T\!\!\! \varphi(T\!\!-\!s)E_t(N^t_s\sqrt {Y_s })ds.
$$
By using Fourier transform techniques, we can evaluate $U$ and its derivatives. Indeed being the Heston model affine,  from  \cite{He} we have
$$
U(t,x,y)= \mathrm e^xP_1(t,x,y)- \mathrm e^\kappa P_2(t,x,y),
$$
where for $j=1,2$, setting  $ \delta_1 =1, \delta_2=0$, we have
\begin{equation}
\label{transf}
\begin{aligned}
P_j(t,x,y)&=\frac 12 +\frac 1 \pi\int_0^{+\infty}\textrm {Re}\Big [ \frac{\mathrm e^{-i \zeta\kappa} f_j(t,x,y,\zeta)}{i\zeta} \Big ]d\zeta\\
f_j(t,x,y,\zeta)&= \mathrm e^{C_j(t,T, \zeta)+ D_j(t,T, \zeta)y+i\zeta  x}\\
C_j(t,T, \zeta)&= \frac{k\theta}{c^2}\Big \{[k-\eta c( \delta_j+ i \zeta)+d_j](T-t)- 2 \ln \Big (\frac {1- g_j\mathrm e^{d_j(T-t)}}{1- g_j}\Big )\Big \},\\
D_j(t,T, \zeta)&=g_j\Big (\frac{1- \mathrm e^{d_j(T-t)}} {1- g_j\mathrm e^{d_j(T-t)}}\Big )\\
g_j&= \frac{k-\eta c( \delta_j+ i \zeta)+d_j}{k-\eta c( \delta_j + i \zeta)-d_j}\quad
d_j=\sqrt{(k-\eta c( \delta_j+ i \zeta)^2 -c^2[ (-1)^{j-1}-\zeta]\zeta}.
\end{aligned}
\end{equation}
and, passing the derivative under the  integral sign,  we may conclude that
\begin{equation}
\begin{aligned}
\partial_x U(t,x,y)&= \mathrm e^x(P_1 +\partial_x P_1)-\mathrm e^\kappa\partial_x P_2,\\
\partial_x P_j(t,x,y)&=\frac 1 \pi\int_0^{+\infty}\textrm {Re}\Big [ \mathrm e^{i \psi \kappa} f_j(t,x,y,\zeta)\Big ]d\zeta,\\
\partial_y U(t,x,y)&=
[\mathrm e^x\partial_y P_1-\mathrm e^\kappa\partial_y P_2](t,x,y),\\
\partial_y P_j(t,x,y)&=\frac 1 \pi\int_0^{+\infty}\textrm {Re}\Big [ \frac{\mathrm e^{-i \zeta \kappa}D_j(t,T, \zeta) f_j(t,x,y,\zeta)}{i\zeta}\Big ]d\zeta.
\end{aligned}
\end{equation}
It remains to compute $\E_t(N^t_s\sqrt {Y_s })$.
Indeed it does  not  exists an explicit expression of $Y$, so we need a manageable  approximation of this process to be able to compute the previous expectation. Since our choice of parameters ($ c^2< 2k \theta$) guarantees the strict positivity of  $Y$, we suggest to approximate it by a geometric Brownian motion
$$
Y_s\approx Y_t\, \mathrm e^{\int_t^s [\gamma_1(u) -\frac{\gamma^2_2(u)}2]du+\int_t^T\gamma_2(u) dB^2_u}
$$
where the  deterministic functions $\gamma_1$ and $\gamma_2$ are chosen appropriately.
To determine such functions we  perform a moment matching between the two random variables, because we can compute explicitly  the first two moments of $Y_s$ given  $Y_t=y$
\begin{equation}
\label{moment1}
\begin{aligned}
m_1(t,s)&:=
\E_t(Y_s)=\theta + (y-\theta ) \mathrm e^{-k(s-t)}\\
m_2(t,s)\!&:=\!
\E_t(Y_s^2)\!=\![(y-\theta)^2\!-\frac {c^2}k(y-\frac \theta 2)]\mathrm e^{-2k(s-t)} \! + (y\!-\theta )(2\theta\!+ \frac {c^2}k) \mathrm e^{-k(s-t)}\!+ \theta(\theta\!+\!\frac {c^2}{2k}),
\end{aligned}
\end{equation}
while on the other hand we have
$$
\begin{aligned}
\label{moment3}
\E_t(y\mathrm e^{\int_t^s [\gamma_1(u) -\frac{\gamma^2_2(u)}2]du+\int_t^T\gamma_2(u) dB^2_u})&= y \mathrm e^{\int_t^s \gamma_1(u) du}\\
\label{moment4}
\E_t(y^2 \mathrm e^{\int_t^s [2\gamma_1(u) -\gamma^2_2(u)]du+\int_t^T2\gamma_2(u) dB^2_u})&=
y^2\mathrm e^{\int_t^s [2\gamma_1(u) +\gamma^2_2(u)]du}.
\end{aligned}
$$
Matching the expectations,  we obtain the system
$$
\begin{cases}
&\!\!\!\!\!y \mathrm e^{\int_t^s \gamma_1(u) du}=m_1(t,s)\\
&\!\!\!\!\!y^2\mathrm e^{\int_t^s [2\gamma_1(u) +\gamma^2_2(u)]du}\!= m_2(t,s)
\end{cases}\,\,\Rightarrow\,\,
\begin{cases}
&\!\!\!\!\!\displaystyle\gamma_1(s) =\partial_s \ln \Big [m_1(t,s)\Big ]= \frac {\partial_s m_1(t,s)}{m_1(t,s)}\\
&\!\!\!\!\!\displaystyle\gamma^2_2(s)=  \partial_s\ln \!\Big [m_2(t,s)\Big ]\!-\!2\gamma_1(s)=\frac {\partial_s m_2(t,s)}{m_2(t,s)}\!-\!2\gamma_1(s),
\end{cases}
$$
explicitly solved by
$$
\!\!\!\!\!\!\!\!\!
\begin{cases}
\!\!\!\!\!\!&\displaystyle\gamma_1(s) =-\frac{k(y-\theta)\mathrm e^{-k(s-t)}}{(y-\theta)\mathrm e^{-k(s-t)}+\theta}\\
\!\!\!\!\!\!&\displaystyle\gamma^2_2(s)=\frac{-2k [(y-\theta)^2-\frac {c^2}k(y-\frac \theta 2)]\mathrm e^{-2k(s-t)}  -k (y-\theta )(2\theta+ \frac {c^2}k) \mathrm e^{-k(s-t)}}{[(y-\theta)^2-\frac {c^2}k(y-\frac \theta 2)]\mathrm e^{-2k(s-t)}  + (y-\theta )(2\theta+ \frac {c^2}k) \mathrm e^{-k(s-t)}+ \theta(\theta+\frac {c^2}{2k})}-2\gamma_1(s).
\end{cases}
$$
Consequently, employing this approximation we have
\begin{equation}
\begin{aligned}
\label{approx3}
\sqrt{Y_s}&\approx \sqrt{y} \mathrm e^{\int_t^s [\frac{\gamma_1(u)}2 -\frac{\gamma^2_2(u)}4]du+\int_t^s\frac{\gamma_2(u)}2 dB^2_u}
\\
\E_t(\sqrt{Y_s})&\approx  \sqrt{y} \mathrm e^{\int_t^s [\frac{\gamma_1(u)}2 -\frac{\gamma^2_2(u)}8]du}
\end{aligned}
\end{equation}
and when evaluating  $\E_t(N^t_s\sqrt {Y_s })$, we may approximate it by the same formula as the first of \eqref{finalHW1}, choosing
$$
b(u)= [\frac{\gamma_1(u)}2 -\frac{\gamma^2_2(u)}8], \qquad c(u)= \frac{\gamma_2(u)}2.
$$

\subsection{The CIR intensity model}

\label{subsec3.4}

In this section we consider a CIR dynamics for the intensity process, the advantage being that, under  the Feller condition, it guarantees the positivity of the process preserving the  mean reversion property.
Hence we choose
\begin{equation}
\label{CIRint}
d\lam_s= q(\mu-\lam_s) ds + \sigma\sqrt{\lam_s}dB_s^3, \qquad \lambda_t = \lambda>0, \quad s>t
\end{equation}
given that $\sigma^2< 2q\mu$, with $\sigma, \mu, q>0$.

We know that, by Fourier inversion, the martingale $\displaystyle N^t_s= \e^{-\int_t^s\lambda_u du} \e^{ \varphi(T-s)\lambda_s + \psi(T-s)}$ has an explicit expression (see\cite{LL}) with
$$
\varphi(T-s)=- \frac { 2( \e^{p(T-s)}-1)}{ p-q+(p+q)\e^{p(T-s)} },\qquad
\psi(T-s)=-\frac {2q\mu}{\sigma^2}\ln \big [ \frac { 2p\e^{(p+q)(T-s)} }{p-q+(p+q)\e^{p(T-s)}}\Big ],
$$
with $p^2= q^2+2\sigma^2$ and it  has dynamics
$dN^t_s= -\sigma \varphi(T-s) N^t_s\sqrt{\lambda_s}dB^3_s.$

\subsubsection{SABR - C I}

\label{subsec3.5}

Combining the above with the SABR model \eqref{SABR2}, we obtain that
\begin{equation}
\label{covar2}
\begin{aligned}
d\langle N^t, X\rangle_s&= -\sigma \rho\varphi(T-s) N^t_s \sqrt {\lambda_s} Y_s \mathrm e^{-(1-\gamma)X_s} ds\\
d\langle N^t, Y\rangle_s&= -\sigma \nu c \varphi(T-s) N^t_s \sqrt {\lambda_s} Y_s ds,
\end{aligned}
\end{equation}
so that  \eqref{cva1bis} may be approximated by
\begin{equation}
\label{CIRSABR1}
CVA(t) \approx (1\!- \!N^t_T) U(t,x,y)\!+\!\sigma\Big [ \rho\e^{-(1-\gamma) x} \partial_x U+\nu  c \partial_y U\Big ](t,x,y)\!\int_t^T\!\!\!\varphi (T\!-\!s)\E_t(N^t_s \sqrt{\lambda_s } Y_s)ds.
\end{equation}
It remains to compute $\E_t(N^t_s \sqrt{\lambda_s } Y_s)$. Due to the empirical experience that the dependence between default and stochastic volatility is rather weak, practioners often consider them as independent, hence  a rough, but handy way, to approximate this factor might be
$$
\E_t(N^t_s \sqrt{\lambda_s } Y_s)=\E_t(N^t_s \sqrt{\lambda_s })\E_t( Y_s)=y\E_t(N^t_s \sqrt{\lambda_s })
$$
being $Y$ a martingale. It remains to compute $\E_t(N^t_s \sqrt{\lambda_s })$ and we remark that applying the integration by parts  formula, we may write
\begin{equation}
\label{lamN}
d(\sqrt{\lambda_s}N^t_s )=\frac 12 \sqrt{\lambda_s}N^t_s\Big [ \frac {4q\mu - \sigma^2}{4\lambda_s}- ( q+ \sigma^2 \varphi(T\!-\!s))\Big ] ds +\sigma N^t_s\Big [\frac 12-  \varphi(T-\!s)\lambda_s\Big ]dB^3_s
\end{equation}
whence, by considering that the martingale part gives null contribution, we have
$$
\E_t(\sqrt{\lambda_s}N^t_s )=\sqrt \lambda N^t_t+ \int_t^s \frac 12\E_t\Big (\sqrt{\lambda_u}N^t_u\Big [ \frac {4q\mu - \sigma^2}{4\lambda_u}- (q+ \sigma^2 \varphi(T-u) )\Big ] \Big ) du
$$
and we decide to approximate this expectation by freezing the $\frac 1{\sqrt\lambda_s}$ factor at the initial value  and solving the resulting ordinary differential  equation we obtain
\begin{equation}
\label{Nsqlam}
\begin{aligned}
\E_t(\sqrt{\lambda_s}N^t_s )&\approx N^t_t\Big [ \e^{-\int_t^s \alpha(u) du}\sqrt \lambda +
\frac {4q\mu-\sigma^2}{8\sqrt\lam} \int_t^s
\e^{-\int_u^s \alpha_r dr }du\Big ]\\
\textrm{ where}\quad
\alpha(u)&=\frac {q + \sigma^2 \varphi(T-u)}4.
\end{aligned}
\end{equation}

\subsubsection{Hull and White - C I}
\label{subsec3.6}

In this case we have
\begin{eqnarray*}
d\langle X, N^t\rangle_s&=&- \rho\sigma \varphi(T-s) N^t_s\sqrt{\lambda_s}Y_sds\\
d\langle Y, N^t\rangle_s&=&-2c(s) \sigma \varphi(T-s) N^t_s\sqrt{\lambda_s}Y_s ds,
\end{eqnarray*}
then the first order approximation becomes
\begin{equation}
\label{cvaHW}
\begin{aligned}
CVA(t)\approx&
(1\!-\!N^t_t )U(t,x,y)+\rho \sigma  \partial_x U(s,x,y) \!\int_t^T\!\!\! \varphi(T-s) \E_t( N^t_s\sqrt{\lambda_s}Y_s) ds\\
&+  \nu\sigma \partial_{y} U(s,x,y)\!\int_t^T\!\!\! \varphi(T-s) c(s) \E_t( N^t_s\sqrt{\lambda_s}Y_s )ds.
\end{aligned}
\end{equation}
As before, we approximate  $\E_t (N^t_s\sqrt{\lambda_s}\sqrt {Y_s } )$ by $\E_t( N^t_s\sqrt{\lambda_s})\E_t(\sqrt {Y_s} )$, with $\E_t( N^t_s\sqrt{\lambda_s})$ approximated as in  \eqref{Nsqlam} and
$$
\E_t(Y_s)=y \e^{\int_t^s [b(u)- \frac {c^2(u)}2] du}.
$$

\subsubsection{Heston - C I}

\label{subsec3.7}

In this case we have
\begin{eqnarray*}
d\langle X, N^t\rangle_s&=&- \rho\sigma \varphi(T-s) N^t_s\sqrt{\lambda_s}\sqrt {Y_s }ds\\
d\langle Y, N^t\rangle_s&=&-c \nu\sigma \varphi(T-s) N^t_s\sqrt{\lambda_s}\sqrt{Y_s}ds
\end{eqnarray*}
which implies the approximation formula
\begin{equation}
\label{cvaHC}
CVA(t)\approx
(1\!-\!N^t_t )U(t,x,y)+\sigma [ \rho \partial_x U(s,x,y) + \!c \nu \partial_{y} U(s,x,y)]\!\int_t^T\!\!\! \varphi(T-s) \E_t\Big [ N^t_s\sqrt{\lambda_s}\sqrt {Y_s }\Big ]ds
\end{equation}
As before, we approximate  $\E_t (N^t_s\sqrt{\lambda_s}\sqrt {Y_s } )$ by $\E_t( N^t_s\sqrt{\lambda_s})\E_t(\sqrt {Y_s} )$, with $\E_t( N^t_s\sqrt{\lambda_s})$ approximated as in  \eqref{Nsqlam} and
the factor  $\E_t\Big [\sqrt{Y_s}\Big ]$ can be computed by lognormal approximation exactly as before by \eqref{approx3}.

\section{The second order approximation}

\label{sec4}

When choosing a CIR model for the  intensity process, as shown in \cite{BRH18}, even in the simpler Black and Scholes model the effect of the correlation parameter between asset's price and intensity in the CVA evaluation is more marked and it requires an approximation that accounts for terms of order $ \rho^2$. This is justified also by the Monte Carlo simulations run in this context. Instead, the effect due to  correlation between intensity and volatility always seems  to be  quite irrelevant.

To capture this behavior, we are going to employ the second order representation formula \eqref{cva1.3}, which leads to a  second order expansion  in $\rho$. Indeed the first order approximation applied in the previous section, due to the freezing of the terms containing $X$, will work efficiently only when the dependence upon $\rho$ is roughly linear.
The computations needed to exploit  representation formula \eqref{cva1.3} are rather lengthy, hence, for the sake of exposition we are going to specify them only for the more commonly used SABR and  Heston models coupled with CIR and under the assumption that $\nu =0$, as we remarked that the behaviour in this parameter is usually very well captured by the first order approximation.

For $\nu=0$, $ \langle Y, \lam\rangle_s=0$ and consequently $\ds H^y_s=N^t_s \frac{d\langle Y, \lam\rangle_s}{ds}=0$, so
 \eqref{cva1.3}  becomes
\begin{equation}
\begin{aligned}
&\textrm{CVA}(t)=(1\!-\!N^t_t )U(t,x,y)+   \partial_x U(t,x,y)H^x_t  \int_t^T\!\!\! \varphi(T\!-s)ds \\
&+(1-\gamma)\int_t^T\!\!\! \varphi(T\!-s)\!\int_t^s\!\!\!\!\!  \E_t\Big [H^x_u   \cL^{1,x}  (\partial _xU )(u,X_u,Y_u)\Big ]duds\\
&+\int_t^T\!\!\! \varphi(T-s)\E_t\Big [\int_t^s\!\!\!\! \partial_x U(u,X_u,Y_u) dH^x_u\Big ] ds\\
&+\int_t^T\!\!\! \varphi(T-s)\E_t\Big [\int_t^s\!\!\!\partial_{xx}U (u, X_u, Y_u) d\langle H^x, X\rangle_u+\int_t^s\!\!\!\partial_{xy}U (u, X_u, Y_u) d\langle H^x, Y\rangle_u\Big ] ds.
\end{aligned}
\end{equation}
where
$\displaystyle
H^x_s= N^t_s  \frac{d\langle X, \lam\rangle_s}{ds}.
$

\subsection{SABR - C II}

\label{sunbsec4.1}

In the case of SABR model we have
$$
\begin{aligned}
H^x_s=\rho \sigma N^t_s \sqrt{\lam_s} Y_s \e^{-(1-\gamma)X_s},\quad
\cL^{1,x} =y^2 \mathrm e^{-2(1-\gamma)x}(\partial_x -1) +\eta c y^2  \e^{-(1-\gamma)x}\partial_y
\end{aligned}
$$
so that we have
$$
\begin{aligned}
&\textrm{CVA}(t)=(1\!-\!N^t_t )U(t,x,y)  + \rho \sigma \Big \{ \sqrt{\lam}y \e^{-(1-\gamma)x}  N^t_t \partial_x U(t,x,y) \int_t^T\!\!\! \varphi(T\!-s)ds \\
&+ (1\!-\!\gamma)\!\!\int_t^T\!\!\!\! \varphi(T\!\!-\!s)\!\int_t^s\!\!\!\! \E_t\Big [ N^t_u \sqrt{\lam_u} Y^3_u \e^{-2(1-\!\gamma)X_u}[  \e^{-(1-\!\gamma)X_u} (\partial_{xx}\! -\partial_x) +\eta c  \partial_{xy} ]U (u,\!X_u,\!Y_u)\Big ]duds\\
&+\int_t^T\!\!\! \varphi(T-s)\E_t\Big [\int_t^s\!\!\!\! \partial_x U(u,X_u,Y_u) d(N^t \sqrt{\lam} Y \e^{-(1-\gamma)X})_u\Big ] ds\\
&+\int_t^T\!\!\! \varphi(T-s)\E_t\Big [\int_t^s\!\!\!\partial_{xx}U (u, X_u, Y_u) d\langle N^t \sqrt{\lam} Y \e^{-(1-\gamma)X}, X\rangle_u\\
&+\int_t^s\!\!\!\partial_{xy}U (u, X_u, Y_u) d\langle N^t \sqrt{\lam} Y \e^{-(1-\gamma)X}, Y\rangle_u\Big ] ds\Big \}
\end{aligned}
$$
Recalling \eqref{lamN} and
\begin{equation}
\label{gamX}
d\e^{-(1-\gamma)X_s}  = \frac{(1-\gamma)(2-\gamma)}2 Y^2_s \e^{-3(1-\gamma)X_s} ds
- (1-\gamma)Y_s\e^{-2(1-\gamma)X_s} dB^1_s
\end{equation}
and applying the multidimensional It\^o's formula with
$f(w,y,z)=wyz$ respectively to the processes $W_s=N^t_s\sqrt{\lambda_s}$, $Y_s$, $Z_s= \mathrm e^{-(1-\gamma)X_s} $, we have that
$$
\begin{aligned}
\frac {dH^x_u}{\rho\sigma}&= \Big\{N^t_u\sqrt{\lam_u}Y_u\e^{-(1-\gamma)X_u}\Big [
\frac {4q\mu - \sigma^2}{8\lambda_u}- \frac{q+ \sigma^2 \varphi(T\!-\!u)}2\ +\frac{(1-\gamma)(2-\gamma)}2 Y^2_u\e^{-2(1-\gamma)X_u}\Big ]  \\
&-(1-\gamma) N^t_u Y^2_u\e^{-2(1-\gamma)X_u}\Big [ \rho \sigma\Big (\frac 12 - \varphi(t-u)\lam_u\Big  )+c\eta \sqrt{\lam_u}\Big] \Big \}du\\
&-(1-\gamma) N^t_u  \sqrt{\lam_u}Y^2_u\e^{-2(1-\gamma)X_u}dB^1_u+c N^t_u\sqrt{\lam_u}Y_u\e^{-(1-\gamma)X_u}dB^2_u\\
&+\sigma  N^t_u Y_u\e^{-(1-\gamma)X_u} \Big [\frac 12-  \varphi(T-\!u)\lambda_u\Big ]dB^3_u
\end{aligned}
$$
whence
$$
\begin{aligned}
d\langle N^t \sqrt{\lam} Y \e^{-(1-\gamma)X}, X\rangle_u&=  N^t_u  \sqrt{\lam_u}Y^2_u\e^{-2(1-\gamma)X_u}[ c\eta-(1-\gamma)Y_u\e^{-(1-\gamma)X_u}]\\
&+ \rho\sigma
 N^t_u  Y^2_u\e^{-2(1-\gamma)X_u} \Big [\frac 12- \varphi(T-u)\Big ]du\\
d\langle N^t \sqrt{\lam} Y \e^{-(1-\gamma)X}, Y\rangle_u&= N^t_u  \sqrt{\lam_u}Y^2_u\e^{-(1-\gamma)X_u}[ c^2 -c\eta(1-\gamma)Y_u\e^{-(1-\gamma)X_u}]du
\end{aligned}
$$
So freezing at the initial time the processes $ \partial_x U(s,X_s,Y_s),  \partial^2_{xx} U(s,X_s,Y_s),  \partial^2_{xy} U(s,X_s,Y_s)$, $\frac 1{\sqrt\lambda_s} $ and $ \mathrm e^{-(1-\gamma)X_s} $, keeping in mind that the martingale parts give no contributions,
we arrive at (after some  lengthy calculations) the following approximation formula

\begin{equation}
\label{cvSABRfin}
CVA(t)
\approx CVA^{(0)}(t)+\rho\sigma CVA^{(1)}(t)+ (\rho\sigma)^2 CVA^{(2)}(t),
\end{equation}
where, setting
$$
F^i_u(x,y)=\e^{-i(1-\gamma)x}\E(Y^i_u)= \e^{-i(1-\gamma)x}y^i \e^{\frac{(i-1)c^2}2(u-t)}, \, i\in \mathbb N\quad \varphi_1(T\!-\!u)=\displaystyle  \frac {4q\mu\!-\!\sigma^2}{8\sqrt\lam}-\frac{q+\sigma^2\varphi(T\!\!-\! u)}2,
$$
we have
$$
\begin{aligned}
CVA^{(0)}(t)&= (1-N^t_t )U(t,x,y) \\
CVA^{(1)}(t)&= \partial_x U(t,x,y)F^1_t(x,y)\!\! \int_t^T\!\!\!\!  \varphi(T\!\!-s)\Big \{N^t_t\sqrt\lam +\!\int_t^s\!\!
\varphi_1(T\!-\!u) \E_t( N^t_u\sqrt{\lambda_u} ) \Big \}duds\\
&-(1-\gamma)  \partial_x U(t,x,y) \int_t^T\!\!\!\!  \varphi(T\!-s)  \int_t^s\E_t( N^t_u\sqrt{\lambda_u} ) \Big \{   c\eta  F^2_u(x,y)+\frac \gamma 2 F^3_u(x,y)\Big\}duds\\
&+ \eta  c\partial_{xx}U(t,x,y)  \int_t^T\!\!\!\varphi(T\!\!-s) \int_t^s \E_t( N^t_u\sqrt{\lambda_u} ) F^2_u(x,y)duds\\
&+ c^2\partial_{xy}U(t,x,y) \int_t^T\!\!\!\varphi(T\!\!-s)   \int_t^s\ \E_t( N^t_u\sqrt{\lambda_u})  F^2_u(x,y)\e^{(1-\gamma)x}dr ds\Big\}\\
 CVA^{(2)}(t)&= [\partial_{xx}U \!\!- (1\!-\!\gamma) \partial_x U](t,x,y) \!\!\int_t^T\!\!\!\varphi(T\!\!-s)  \!\!  \int_t^s\!\! F^2_u(x,y)\Big [\frac {N^t_t}2 - \varphi(T\!\!-\!u)\E_t(  N^t_u\lambda_u)  \Big  ]duds,
\end{aligned}
$$
it only remains to evaluate
\begin{equation}
\label{Nlam0}
\begin{aligned}
&\E_t(N^t_u\lambda_u)=N^t_t\lambda+ q\mu\int_t^u \E_t (N^t_r)dr-  \int_t^u[q+\sigma^2 \varphi(T-r)]\E_t(N^t_r\lambda_r)dr\\
\Rightarrow\quad  &\E_t(N^t_u\lambda_u)=N^t_t\e^{-  \int_t^u[q+\sigma^2 \varphi(T-r)]dr}\Big \{\lambda+ q\mu\int_t^u \e^{  \int_t^r[q+\sigma^2 \varphi(T-v)]dv}dr\Big \}
\end{aligned}
\end{equation}
and the integrand
$\E(N^t_s \sqrt{\lambda_s })$, to be done as in  \eqref{Nsqlam}.

\subsubsection{Heston - C II}

\label{subsec4.2}

In this case we have $\gamma =1$  and  $H^x_s= \rho \sigma N^t_s \sqrt{\lam_s}\sqrt{Y_s}$, so that we have
$$
\begin{aligned}
&\textrm{CVA}(t)=(1\!-\!N^t_t )U(t,x,y)+   \rho \sigma \partial_x U(t,x,y)N^t_t \sqrt{\lam y} \int_t^T\!\!\! \varphi(T\!-s)ds \\
&+ \rho \sigma\int_t^T\!\!\! \varphi(T\!\!-s)\E_t\Big [\int_t^s\!\!\!\! \partial_x U(u,X_u,Y_u) d( N^t \sqrt{\lam}\sqrt{Y})_u\Big ] ds\\
&+\! \rho \sigma\!\!\!\int_t^T\!\!\! \!\varphi(T\!\!-\!s)\E_t\Big [\!\!\int_t^s\!\!\! \Big ( \partial_{xx}U (u, X_u, Y_u) (d\langle N^t \sqrt{\lam}\sqrt{Y}, X\rangle_u\!
+\partial_{xy}U (u, X_u, Y_u) d\langle N^t \sqrt{\lam}\sqrt{Y}, Y\rangle_u\Big )\Big ] ds.
\end{aligned}
$$
 We recall that the function $\partial_x U$ verifies the same PDE as $U$,
which implies that $ \partial_x U(s.X_s, Y_s)$ is again a martingale verifying
\begin{equation}
\label{dUx}
d  \partial_x U(u,X_u, Y_u)=  \partial^2_{xx} U(u.,X_u, Y_u)\sqrt {Y_s} d B^1_u+
 c \partial^2_{xy} U(u.,X_u, Y_u)\sqrt {Y_u} d B^2_u.
\end{equation}
We have
$$
\begin{aligned}
&d(N^t \sqrt{\lam}\sqrt{Y})_u=\sqrt{Y_u}d(N^t_u\sqrt{\lambda_u})+ N^t_u\sqrt{\lambda_u}d\sqrt{Y_u}\\
=&\Big \{\frac{N^t_u}8\Big [ (4q\mu-\sigma^2)\sqrt{\frac{Y_u}{\lambda_u}}+(4k\theta-c^2)
)\sqrt{\frac{\lambda_s}{Y_u}}\Big ]-\frac {N^t_u}2\sqrt{\lambda_u}\sqrt{Y_u}[q+k+\sigma^2\varphi(T-u)]\Big \} du\\
&+N^t_u\Big [ \sigma\sqrt{Y_u}(\frac 12 -\varphi(T-u) \lambda_u) dB^3_u+\frac c2\sqrt{\lambda_u} dB^2_u\Big ]\\
&d\langle N^t \sqrt{\lam}\sqrt{Y}, X\rangle_u=\Big [\rho\sigma N^t_u Y_u(\frac 12 -\varphi(T-u) \lambda_u)+\frac c2 \eta N^t_u \sqrt{\lambda_u}\sqrt{Y_u} \Big ]du\\
&d\langle N^t \sqrt{\lam}\sqrt{Y}, Y\rangle_u=\frac {c^2}2 N^t_u  \sqrt{\lambda_u} \sqrt{Y_u} du
\end{aligned}
$$
substituting in the above we obtain
$$
\begin{aligned}
&\textrm{CVA}(t)=(1\!-\!N^t_t )U(t,x,y)+   \rho \sigma \partial_x U(t,x,y)N^t_t \sqrt{\lam y} \int_t^T\!\!\! \varphi(T\!-s)ds \\
&+ \rho \sigma\int_t^T\!\!\! \varphi(T\!\!-s)\E_t\Big [\int_t^s\!\!\!\! \partial_x U(u,X_u,Y_u)\Big \{\frac{N^t_u}8\Big [ (4q\mu-\sigma^2)\sqrt{\frac{Y_u}{\lambda_u}}+(4k\theta-c^2)
)\sqrt{\frac{\lambda_s}{Y_u}}\Big ]\\
&\hskip6.5cm -\frac {N^t_u}2\sqrt{\lambda_u}\sqrt{Y_u}[q+k+\sigma^2\varphi(T-u)]\Big \} du\Big ] ds\\
&+\! \rho \sigma\!\!\!\int_t^T\!\!\! \!\varphi(T\!\!-\!s)\!\!\int_t^s\!\!\!  \E_t\Big (\partial_{xx}U (u, X_u, Y_u)\Big [\rho\sigma N^t_u Y_u(\frac 12 -\varphi(T-u) \lambda_u)+\frac c2 \eta N^t_u \sqrt{\lambda_u}\sqrt{Y_u} \Big ]\\
&\hskip2.7cm +\frac {c^2}2\partial_{xy}U (u, X_u, Y_u)  N^t_u  \sqrt{\lambda_u} \sqrt{Y_u} \Big )du ds.
\end{aligned}
$$
As before, by freezing $U$ and its derivatives, $\frac 1{\sqrt{\lam_u}}$ $\frac 1{\sqrt{Y_u}}$ at the initial points and by exploiting the independence between $Y$ and  $\lambda$, this formula may be  approximated by
\begin{equation}
\label{cvafin}
CVA(t)
\approx CVA^{(0)}(t)+\rho\sigma CVA^{(1)}(t)+ (\rho\sigma)^2 CVA^{(2)}(t),
\end{equation}
where
$$
\begin{aligned}
CVA^{(0)}(t)&= (1-N^t_t )U(t,x,y) \\
CVA^{(1)}(t)&=\Big \{  \partial_x U(t,x,y)\Big [N^t_t\int_t^T \!\! \varphi(T-s) \Big(\sqrt{\lambda y} +\frac{4q\mu-\sigma^2}{8\sqrt \lambda}   \int_t^s\E_t(\sqrt{Y_u})du  \Big )ds\\
&+\int_t^T \!\!\! \varphi(T-s) \int_t^s \Big [ \frac{4k\theta- c^2}{8\sqrt y} -  \frac {q+k+\sigma^2\varphi(T-u)}2 \E_t(\sqrt{Y_u})\Big ]\E_t (N^t_u\sqrt{\lambda_u} )du\,  ds\Big ]
\\
&+ \frac {\eta  c}2 \partial_{xx}U(t,x,y)  \int_t^T  \varphi(T-s) \int_t^s\E_t(\sqrt{Y_u}) \E_t( N^t_u\sqrt{\lambda_u}  ) du ds\\
&+\frac {c^2}2\partial_{xy}U(t,x,y)\int_t^T  \varphi(T-s) \int_t^s\E_t(\sqrt{Y_u}) \E_t( N^t_u\sqrt{\lambda_u}  ) du ds\Big\}\\
 CVA^{(2)}(t)&= \partial_{xx}U(t,x,y) \int_t^T  \varphi(T-s) \int_t^s [ \frac {N^t_t}2- \varphi(T-u)\E_t(  N^t_u\lambda_u)   ]\E_t(Y_u)duds.
\end{aligned}
$$
Again we have to approximate the expectations inside the integrals:
$$
\begin{aligned}
& \E_t(Y_u )\,\, \textrm{as in \eqref{moment1}},\quad
\E_t(\sqrt{Y_u} )\,\, \textrm{as in \eqref{approx3}}, \quad \E_t (N^t_u\sqrt{\lambda_u}) \,\,\textrm{ as in \eqref{Nsqlam}}, \quad\E_t(N^t_u\lambda_u)\,\, \textrm{ as in \eqref{Nlam0}}.\\
\end{aligned}
$$

\section{Numerical results}

\label{sec5}
To show the numerical efficiency of our method,
in this section we present  some numerical implementations for the SABR and Heston stochastic volatility models coupled with either a Vasicek or a CIR intensity model. The Hull-White model shares essentially the same behavior, but we are not going to present it, since in this case the efficiency of  our correlation method is less evident due to the additional error introduced by the approximation of the  zero-th term of the expansion. Numerical experiments were implemented in MatLab (version 9.2.0 - R2017a) on an Intel Core i7 2.40GHZ with 8GB RAM. The CVA  was computed for different values of $\rho$ (the underlying-intensity correlation) spanning the interval $(-1, 1)$ and we compare the results of our approximation formulas  with a full Monte Carlo estimation of (\ref{cva0}) with $R=0$, taken as  the benchmark CVA value.  To reduce the variance of these latter estimates, we used the default-free option value as a control variate: the high correlation  between default  free and defaultable option price (up to $99\%$ in our experiments)  reduces the simulation error  by one order of magnitude.

Simulation of the Heston paths were realized through an Euler scheme with full-truncation for the volatility component \cite{LKVD}. For the SABR model, we implemented an exact simulation of the volatility component and a log-Euler scheme for the asset's price. We considered a uniform time grid $0=t_0 < t_1 < \cdots < t_N=T$, with $N = 10^3$ and $M =10^6$ sample paths.

The implementation of our approximation formulas is quite straightforward: since the integrands  resulted generally well-behaved in all the considered cases, the time (iterated) integrals were evaluated by using a trapezoidal rule with step $\Delta t= 10^{-2}$.

In order to obtain consistent results, we fixed the parameters of the two models  to fit approximately the same (given) set of default-free call prices at time $t=0$: $\kappa=1.15$, $\theta=0.04$, $c=0.39$, $\eta=-0.34$ and $y=0.034$ for the Heston model,  $\gamma=.7367$, $c=0.7356$, $\eta=-0.3$ and $y=.5887$ for the SABR. Without loss of generality we took $r=0$. To keep formulas  simple, we always took $\nu=0$ (the volatility-intensity correlation) in the models, since our main interest lied on the interaction between default and asset's price (WWR).
Moreover, extensive numerical simulations confirmed that the contribution due to the terms multiplied by $\nu$ was usually quite modest.

\medskip

\begin{figure}[h]
\begin{center}
\includegraphics[width=12cm,height=6cm]{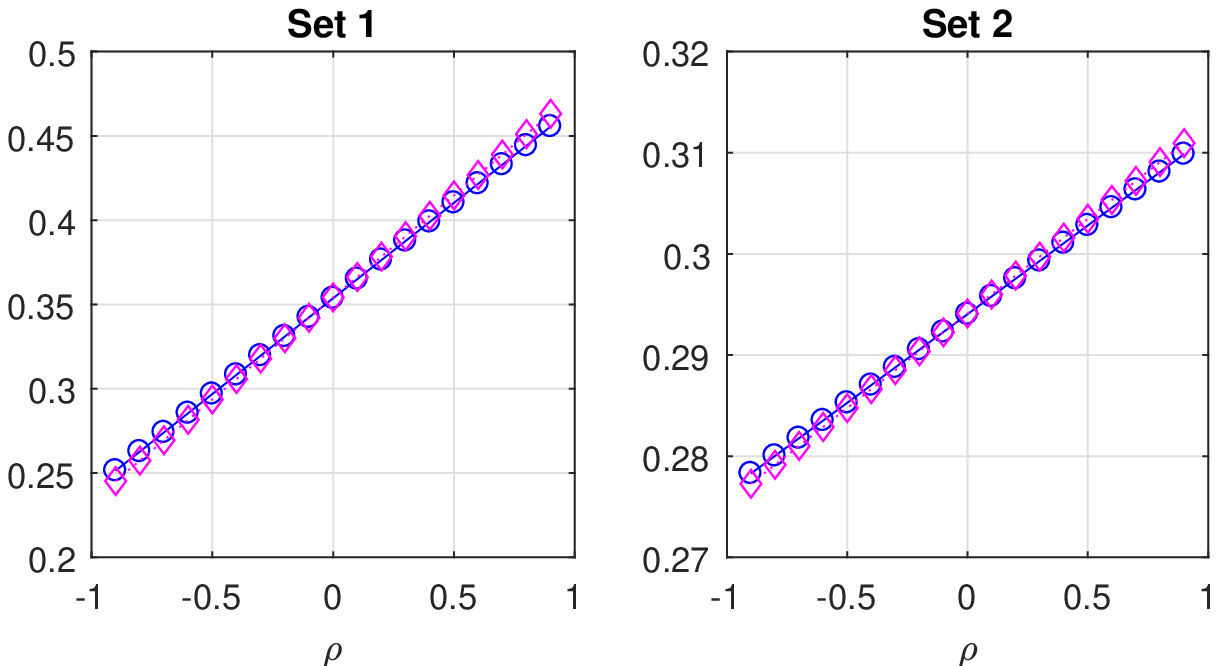}
\end{center}
\caption{CVA profiles for varying $\rho$ in the SABR model with Vasicek intensity of default: $T=1/2$, Monte Carlo (blue-circle), first order approx. (magenta-diamond).} \label{SABR-Vas-05}
\end{figure}

\begin{figure}[h]
\begin{center}
\includegraphics[width=12cm,height=6cm]{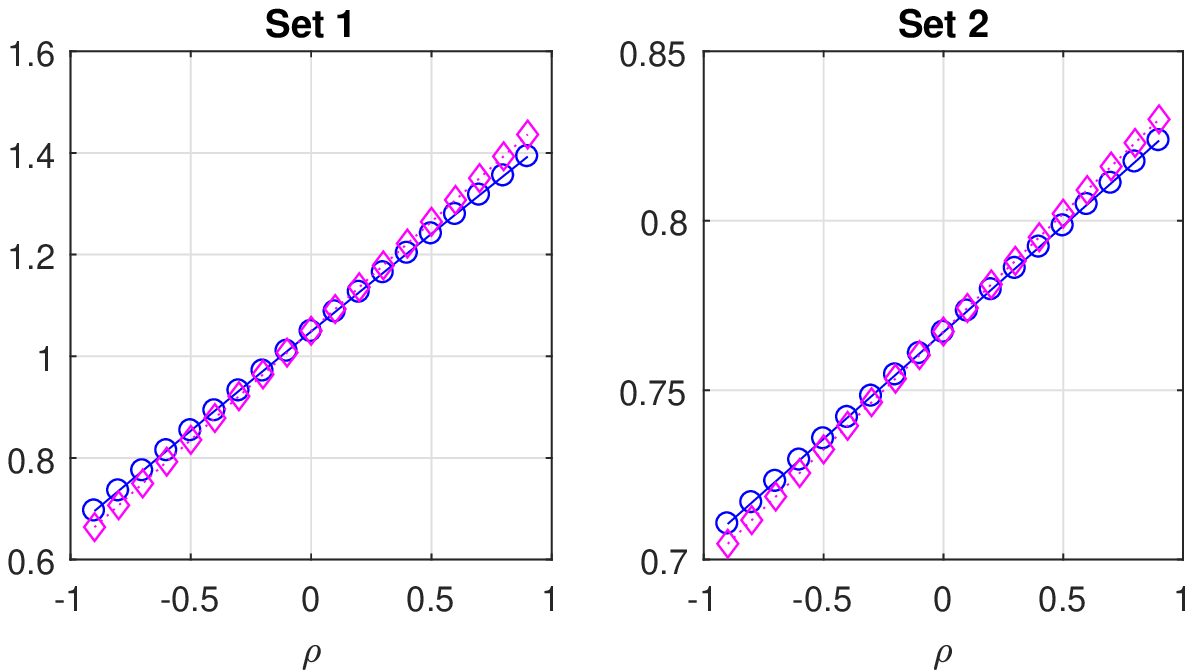}
\end{center}
\caption{CVA profiles for varying $\rho$ in the SABR model with Vasicek intensity of default: $T=1$, Monte Carlo (blue-circle), first order approx. (magenta-diamond).} \label{SABR-Vas-1}
\end{figure}

\begin{figure}[h]
\begin{center}
\includegraphics[width=12cm,height=6cm]{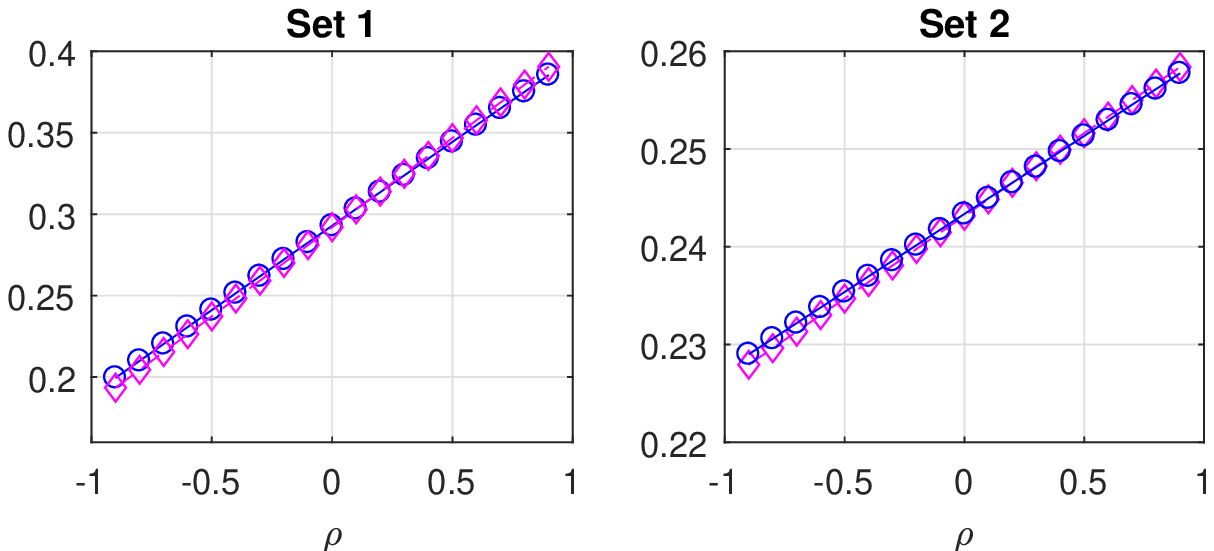}
\end{center}
\caption{CVA profiles for varying $\rho$ in the Heston model  with Vasicek intensity of default: $T=1/2$, Monte Carlo (blue-circle), first order approx. (magenta-diamond).} \label{HESTON-Vas-05}
\end{figure}

\begin{figure}[h]
\begin{center}
\includegraphics[width=12cm,height=6cm]{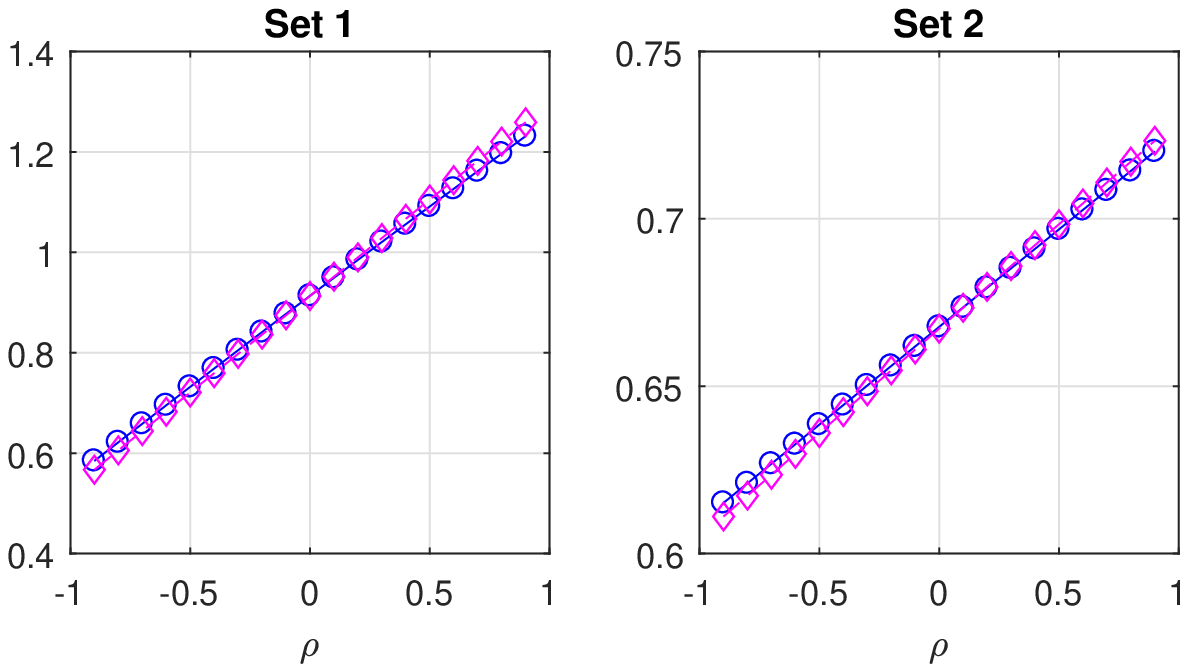}
\end{center}
\caption{CVA profiles for varying $\rho$ in the Heston model with Vasicek intensity of default: $T=1$, Monte Carlo (blue-circle), first order approx. (magenta-diamond).} \label{HESTON-Vas-1}
\end{figure}

As  already  mentioned at the end of section 3 the Vasicek model, even though used in the literature (see e.g. \cite{Fard15}),  has the drawback  of assuming negative values with positive probability. To contain this effect, we arbitrarily chose the Vasicek parameters so that the asymptotic value of the intensity  has a small probability ($<0.1\%$) to become negative (Table (\ref{VasR_param})).
In this context,  since Monte Carlo simulations seemed to exhibit always a linear behavior in  $\rho$, we implemented only the first order approximation, which
performed satisfactorily in both SABR and Heston model (Figs. \ref{SABR-Vas-05}, \ref{SABR-Vas-1}, \ref{HESTON-Vas-05} and \ref{HESTON-Vas-1}).

\begin{table}[t]
\centering
\begin{tabular}{|c|c|c|c|c|}
  \hline
   & $\lambda_0$ & $q$  & $\mu$ &  $\sigma$ \\ \hline
 Set 1  & 0.09 & 0.3 & 0.4 & 0.1\\
 Set 2  & 0.1 & 0.18 & 0.1   & 0.015 \\
  \hline
\end{tabular}
\caption{Parameter sets for the Vasicek default intensity.}
\label{VasR_param}
\end{table}

\medskip

When assuming a  CIR default intensity, we ran our numerics with  three different sets of parameters given in Table (\ref{CIR_param}),  we chose the first  two arbitrarily, while the third  was taken as in  \cite{BV18}, \cite{BRH18} where it was considered consistent with observed patterns of CDS spreads and implied volatilities.

Simulated paths were generated with the same full-truncation scheme used for the volatility component of the Heston model.
\begin{table}[t]
\centering
\begin{tabular}{|c|c|c|c|c|}
  \hline
   & $\lambda$ & $q$  & $\mu$ &  $\sigma$ \\ \hline
 Set 1  & 0.03 & 0.02 & 0.161 & 0.08\\
 Set 2  & 0.05 & 0.09 & 0.2   & 0.1 \\
 Set 3  & 0.01 & 0.8 & 0.02 & 0.2\\
 Set 4  & 0.03 & 0.5 & 0.05 & 0.5\\
  \hline
\end{tabular}
\caption{Parameter sets for the CIR default intensity.}
\label{CIR_param}
\end{table}

\begin{figure}[ht]
\begin{center}
\includegraphics[width=12cm,height=12cm]{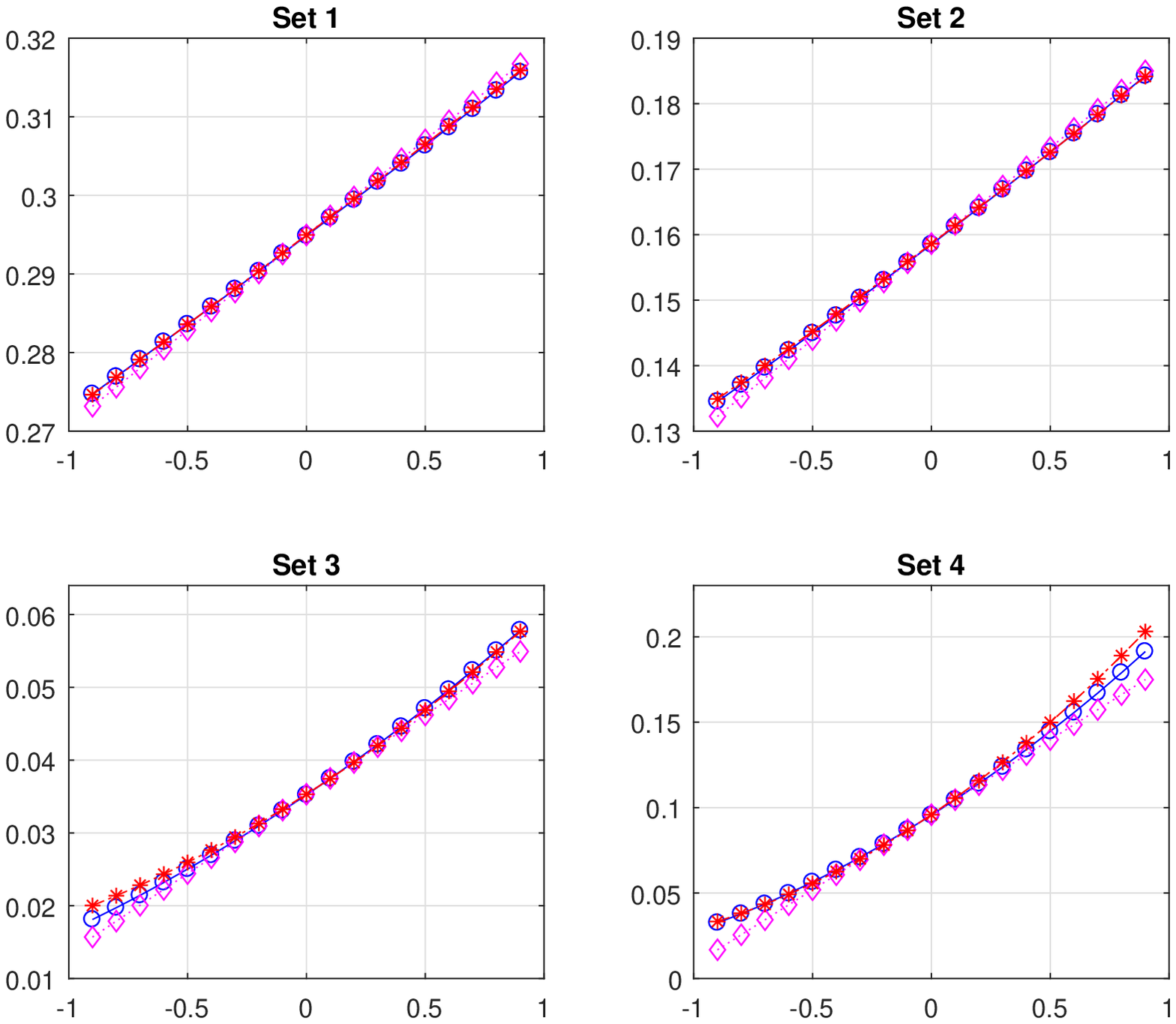}
\end{center}
\caption{CVA profiles for varying $\rho$ in the SABR model with CIR intensity of default: $T=1/2$, Monte Carlo (blue-circle), first order approx. (magenta-diamond), second order approx. (red-star).} \label{SABR-CIR-05}
\end{figure}

\begin{figure}[ht]
\begin{center}
\includegraphics[width=12cm,height=12cm]{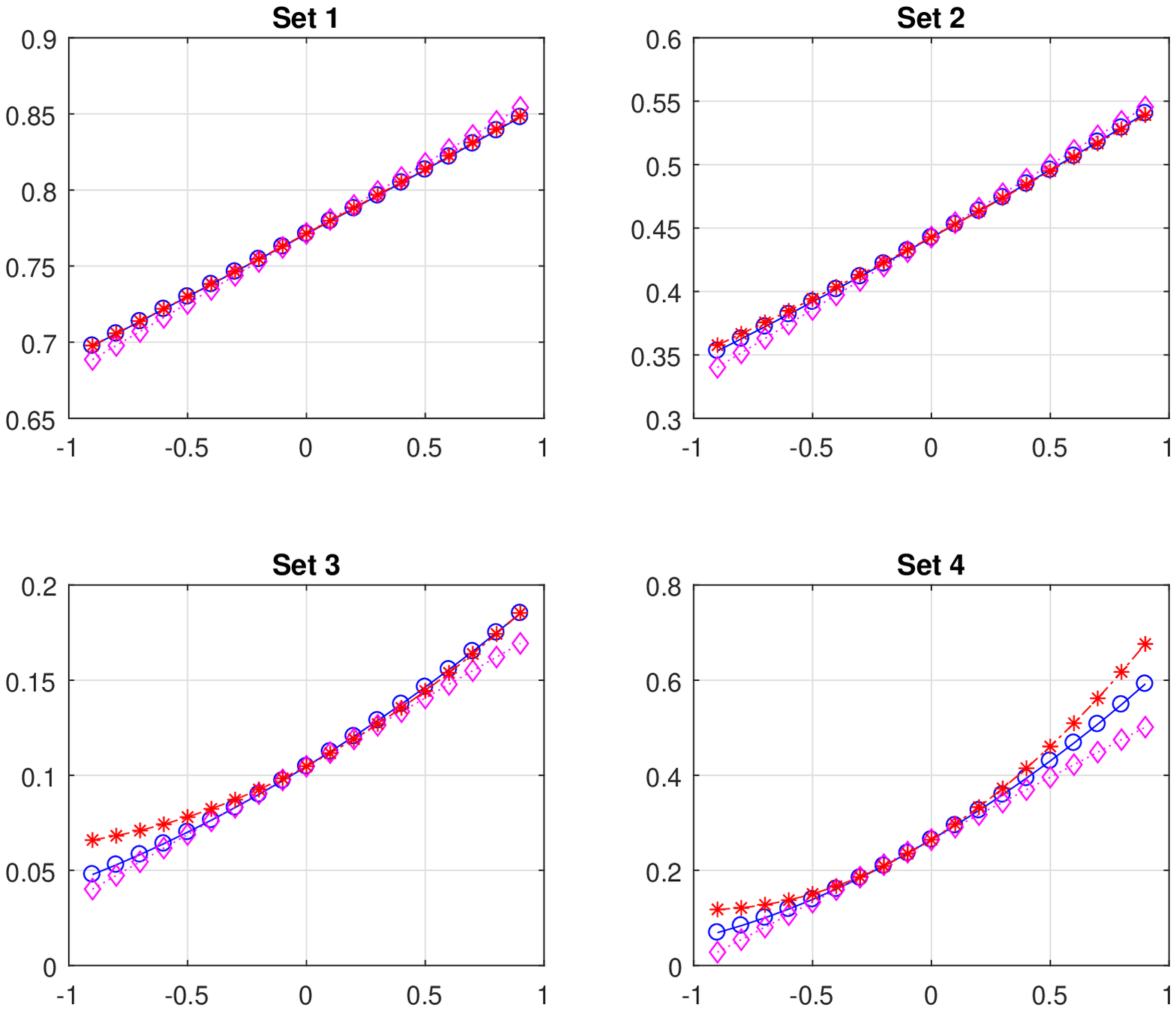}
\end{center}
\caption{CVA profiles for varying $\rho$ in the SABR model with CIR intensity of default: $T=1$, Monte Carlo (blue-circle), first order approx. (magenta-diamond), second order approx. (red-star).}\label{SABR-CIR-1}
\end{figure}

\begin{figure}[ht]
\begin{center}
\includegraphics[width=12cm,height=12cm]{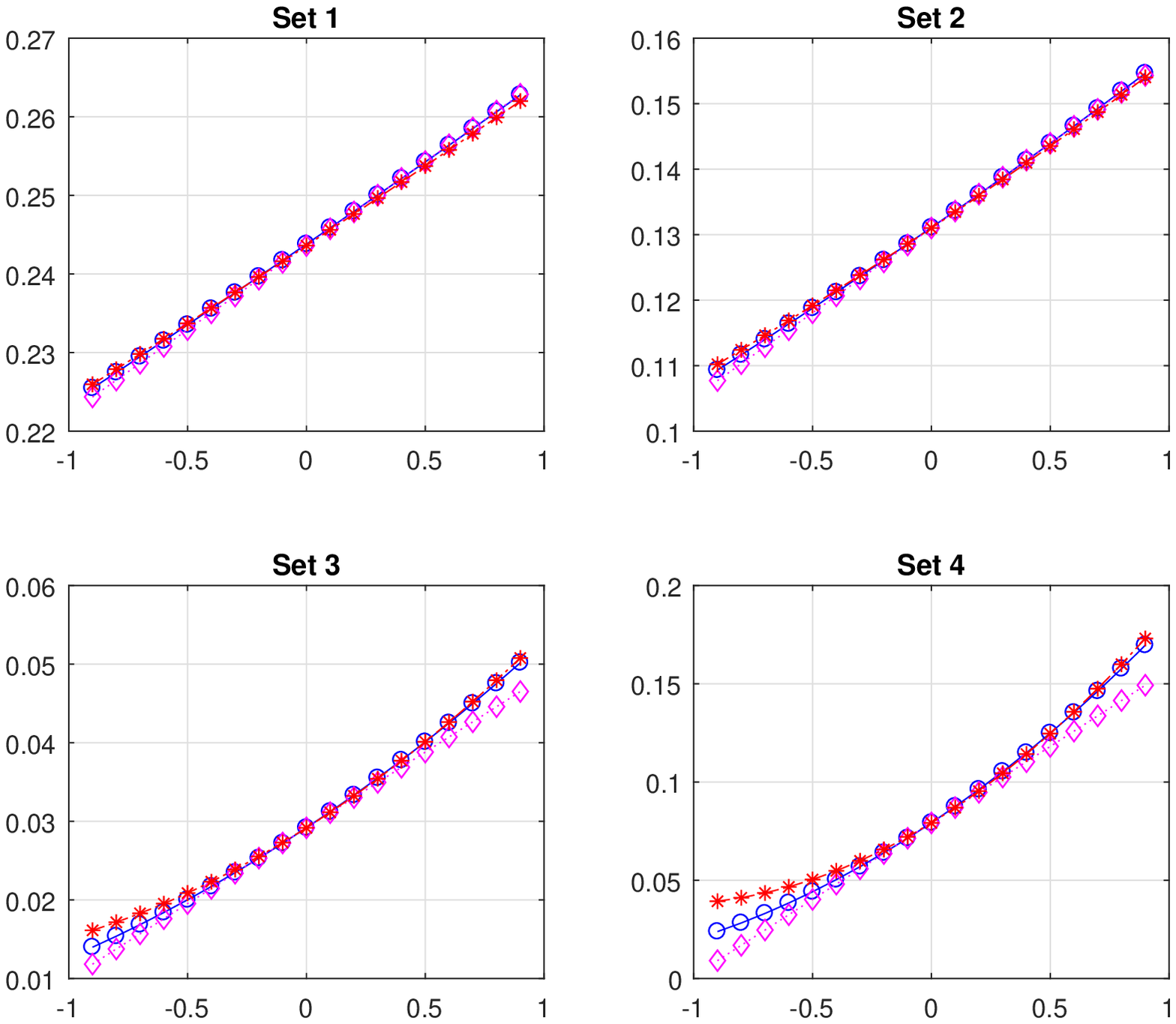}
\end{center}
\caption{CVA profiles for varying $\rho$ in the Heston model with CIR intensity of default: $T=1/2$, Monte Carlo (blue-circle), first order approx. (magenta-diamond), second order approx. (red-star).}\label{HESTON-CIR-05}
\end{figure}

\begin{figure}[ht]
\begin{center}
\includegraphics[width=12cm,height=12cm]{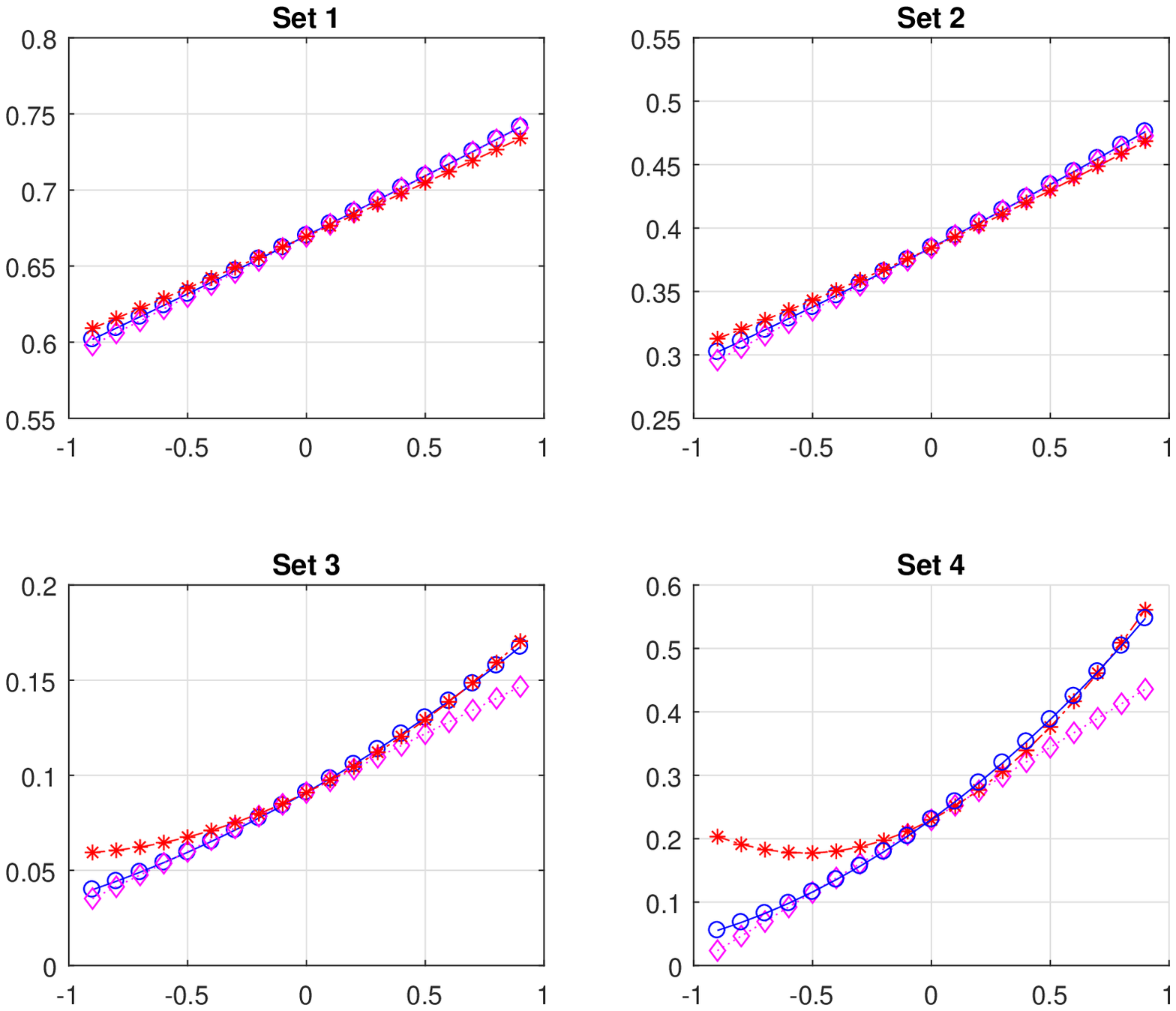}
\end{center}
\caption{CVA profiles for varying $\rho$ in the Heston model with CIR intensity of default: $T=1$, Monte Carlo (blue-circle), first order approx. (magenta-diamond), second order approx. (red-star).}\label{HESTON-CIR-1}
\end{figure}

Figures \ref{SABR-CIR-05}, \ref{SABR-CIR-1} refer to the first and second order formulas (\ref{CIRSABR1}) and (\ref{cvSABRfin})  for the SABR model,  while  figures \ref{HESTON-CIR-05} and \ref{HESTON-CIR-1} correspond to  (\ref{cvaHC}) and (\ref{cvafin}) for the Heston model. We notice that for Sets 1 and 2  both  the first  and the second order approximations  behave quite well for both models and maturities, uniformly in $\rho$, instead when choosing Set 3 the CVA pattern shows a curvature and the second order approximation seems to be more appropriate. Finally we remark that  for  small values of $|\rho|$, approximations are  very satisfactory for all the models, while they always tend to worsen for growing maturities at large values of $|\rho|$ (only for the  negative ones for the Heston model). This effect  might be due to the freezing  we used to produce the approximation formulas.

\begin{figure}[ht]
\begin{center}
\includegraphics[width=12cm,height=6cm]{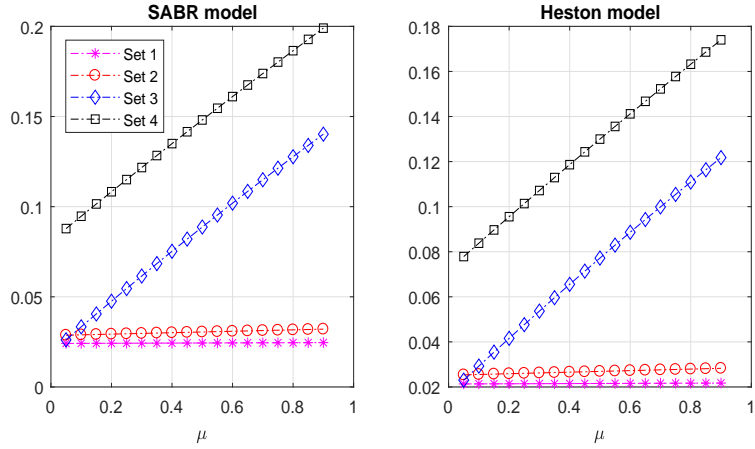}
\end{center}
\caption{$CVA^{(1)}$ as a function of the long term mean  $\mu$ for the two models: here $T=1/2$.}\label{sens_mu}
\end{figure}

\begin{figure}[ht]
\begin{center}
\includegraphics[width=12cm,height=6cm]{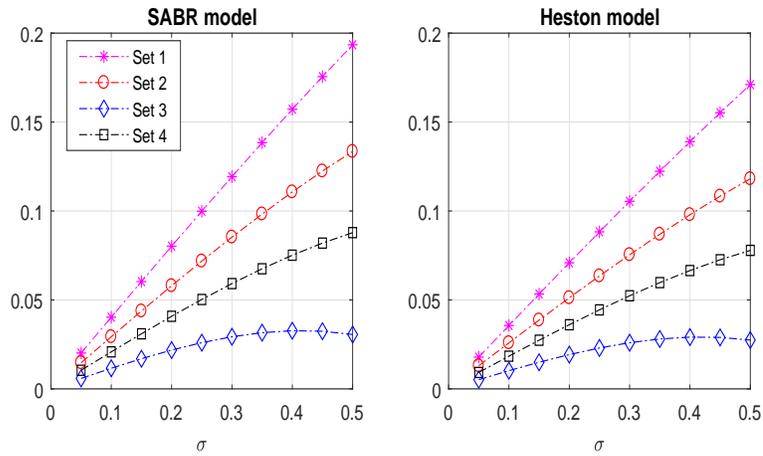}
\end{center}
\caption{$CVA^{(1)}$ as a function of the volatility of the intensity $\sigma$ for the two models: here $T=1/2$.}\label{sens_sigma}
\end{figure}

Finally, we analyzed the impact of the intensity model parameters on the CVA.  Taking the CIR model as an example, Figures \ref{sens_mu} and \ref{sens_sigma} show the impact of the parameters $\sigma$ and $\mu$. In particular, the linear part in (\ref{CIRSABR1}) and (\ref{cvaHC})  represents the contribution to the CVA due to correlation  and we notice that  increasing  $\sigma$ and $\mu$  determines and increase  of  the WWR  effect ($\rho>0$) in all the considered set of parameters.

\smallskip

Set 4 deserves for a separate discussion. Indeed, even though it strongly violates Feller's condition for the intensity, we decided to employ it in the simulations, since it was another   set consistent with the observed patterns of CDS spreads and implied volatilities (see \cite{BRH18}). This is the most troublesome set and even if it shows still acceptable results for short maturities in both SABR and Heston models, so capturing the Right and Wrong Way risks, it becomes unsatisfactory for larger maturities. One might argue that CIR model is not appropriate for this set of parameters as a noticeable probability to assume negative values is a remarkable modeling flaw. Perhaps, in this context, it could be worth to make alternative choices to describe the default intensity and/or to sharpen the approximations used in our representation formulas.  Here we employed the most classical intensity models found in the literature and we postpone to future work more detailed discussion and analysis on the impact of the default model for CVA evaluation.

\newpage

%
%


\end{document}